\documentclass[prl,twocolumn,showpacs,preprintnumbers,amsmath,amssymb,floats,aps,10pt]{revtex4-1}
\pdfoutput=1
\usepackage{graphicx}
\usepackage{amsmath}
\usepackage{pstricks}
\usepackage{animate}
\usepackage{dcolumn}
\usepackage{bm}
\newcommand{\vq}{\mathbf{q}}
\newcommand{\BSCCO}{{Bi$_2$Sr$_2$CaCu$_2$O$_8$ }}

\newcommand\rr{{\mathbf r}}
\newcommand\R{{\mathbf R}}
\begin{document}

\title{Interpretation of scanning tunneling quasiparticle interference and impurity states in cuprates}
\author{A. Kreisel$^{1,4}$, Peayush Choubey$^1$, T. Berlijn$^{2}$,  W. Ku$^3$, B. M. Andersen$^4$ and  P. J. Hirschfeld$^1$}
\affiliation{
$^1$Department of Physics, University of Florida, Gainesville, Florida 32611, USA\\
$^2$Center for Nanophase Materials Sciences and Computer Science and Mathematics Division, Oak Ridge National Laboratory, Oak Ridge, Tennessee 37831, USA\\
$^3$Condensed Matter Physics and Materials Science Department, Brookhaven National Laboratory, Upton, New York 11973, USA\\
$^4$Niels Bohr Institute, University of Copenhagen, DK-2100 Copenhagen, Denmark
}

\date{\today}

\begin{abstract}
We apply a recently developed method combining first principles based Wannier functions with solutions to the Bogoliubov-de Gennes equations to the problem of interpreting STM data in cuprate superconductors.   We show that the observed images of Zn on the surface of  \BSCCO can only be understood by accounting for the tails of the Cu Wannier functions, which include significant weight on apical O sites in neighboring unit cells.
This calculation thus puts earlier crude ``filter" theories on a microscopic foundation and solves a long standing puzzle.
We then study quasiparticle interference phenomena induced by out-of-plane weak potential scatterers, and show how patterns long observed in cuprates can be understood in terms of the interference of Wannier functions above the surface. Our results show excellent agreement with experiment and enable a better understanding of novel phenomena in the cuprates via STM imaging.
\end{abstract}


\maketitle

Scanning tunneling microscopy (STM) methods were applied to cuprates relatively early on, but dramatic improvements in energy and spatial resolution led to a new set of classic discoveries in the early part of the last decade, giving for the first time a truly local picture of the superconducting and pseudogap states at low temperatures\cite{FischerRMP,FujitaJPSJ}.  These measurements revealed gaps that were much more inhomogeneous than had previously been anticipated\cite{Cren01,Pan2001,Kapitulnik1,Lang2002}, exhibited localized impurity resonant states\cite{Yazdani_boundstate,PanZn}, and gave important clues to the nature of competing order\cite{Kapitulnik_checkerboard1,Kivelson_RMP,Davis_checkerboard1,Vershinin04,Davis_checkerboard2}. More recently,
STM has again been at the forefront of studies of inhomogeneities, this time as a real space probe of
intra-unit cell charge ordering visible in the underdoped systems\cite{Fujita14}. 
While a microscopic description of such atomic scale phenomena in superconductors is available in terms of the Bogolibuov de-Gennes equations, such calculations are always performed on a lattice with sites centered on the Cu atoms, and thus do not contain intra-unit cell information.

\begin{figure}
\includegraphics[width=\columnwidth]{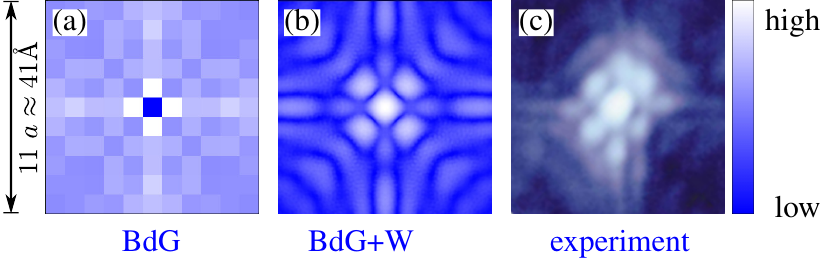}
\caption{(Color online)
(a) Resonant state real-space BdG patterns at $\Omega_0= -3.6\,\text{meV}$ as obtained from conventional BdG calculations in logarithmic scale, (b) xy cut through continuous 3D LDOS $(x,y,z\approx 5\text{\AA};\Omega_0)$  at $\Omega_0=-3.6\,\text{meV}$ showing strong similarity to the measured conductance maps (c) reproduced from Ref. [\onlinecite{PanZn}] rotated to match the orientation in (a) and (b) and cropped to 11$\times$11 elementary cells with the impurity located at the center.}
\label{fig:comparison}
\end{figure}

The simplest example of a problem that can arise because of the deficiencies of theory in this regard is that of the Zn impurity substituting for Cu in  \BSCCO (BSCCO), a cuprate material which cleaves well in vacuum, leaving atomically smooth surfaces ideal for STM. The observation of a spectacularly sharp impurity resonance at the impurity site\cite{Yazdani_boundstate,PanZn,Machida11,Hamidian12} was an important local confirmation of unconventional pairing in the cuprates.
The differential conductance map near the impurity exhibits a cross-shaped real-space conductance map at resonance, as expected for a pointlike potential scatterer in a $d$-wave superconductor, see Fig.~\ref{fig:comparison}(c)
\cite{Note1}
Upon closer examination, however, the pattern deviates from the expected theoretical one on the Cu square lattice  in some important respects\cite{BalatskyRMP,AlloulRMP}.  First, it displays a central maximum on the impurity site, unlike simple models, which have a minimum (Fig. \ref{fig:comparison}(a)).   Second, the longer range intensity tails
are rotated 45 degrees from the nodal directions of the $d$-wave gap, where such long quasiparticle decay lengths are expected\cite{BalatskyRMP}.     There  is still no consensus on the origin of this pattern, which has been discussed in terms of nonlocal Kondo correlations\cite{Polkovnikov01}, postulated extended potentials\cite{Flatte00,Tang02,Tang04}, Andreev phase impurities\cite{Andersen06}, and ``filter effects", which assume that the tunneling process from the surface to the impurity through several insulating layers  involves atomic states in several neighboring unit cells\cite{Ting,Martin}.   So far, these theories have been expressed entirely in terms of phenomenological effective hoppings in the Cu tight-binding model.
First principles calculations for Zn in BSCCO in the normal state\cite{abinitioZn} provide some evidence in support of the filter picture,
but until recently it was not possible to include both superconductivity and the various atomic wave functions extending into the barrier layers responsible for the filter. Nieminen \textit{et al.} investigated the conductance spectrum in the BSCCO system using an analysis based on atomiclike wave functions\cite{Markiewicz}, and showed that for the homogeneous system it could be decomposed in a series of tunneling paths, as postulated by the earlier crude proposals\cite{Ting,Martin}. 
Using this approach one can explain, e.g., the spectral lineshape at high bias voltage, but presently it is unclear how this approach applies to inhomogeneous problems.

The vast amount of STM data on cuprate surfaces have often been distilled using the quasiparticle interference (QPI), or Fourier transform STM spectroscopy technique, one of the most important modern techniques for unraveling the origin of high temperature superconductivity.
This probe is sensitive to the wavelengths of Friedel oscillations caused by disorder, which then, in principle contain information on  the electronic structure of the pure system\cite{Sprunger1997,Hoffman2002}.     These wavelengths manifest themselves in the form of peaks at  wave vectors $\vq(\omega)$, which disperse with STM bias $V=\omega/e$ and represent scattering processes of high probability on the given Fermi surface.  Many attempts have been made to calculate these patterns assuming simple tight-binding band structures, $d$-wave pairing, and methods ranging from single-impurity $T$ matrix\cite{WangLee2003,Capriotti03,Pareg03,zhang03,Andersen03,Nunner06,Nowadnick12} to many-impurity solutions of the BdG equations\cite{lingyin04}. While some similarities between the calculated patterns, the simplified so-called ``octet model"\cite{WangLee2003}, and experiment have been reported, there are always serious discrepancies,  typically related not so much to the positions of peaks but rather their shapes and intensities. 
%

In this paper we revisit these classic unsolved problems using a new method called the BdG-Wannier (BdG+W) approach\cite{Choubey2014}, which combines traditional solutions of the Bogoliubov-de Gennes (BdG) equations with the microscopic Wannier functions obtained from downfolding density functional theory onto a low-energy effective tight-binding Hamiltonian.  We show that the local density of states (LDOS) obtained from the continuum Green's function for a simple strong nonmagnetic impurity bound state in the BSSCO material with a $d$-wave superconducting gap displays excellent agreement with STM conductance maps (Fig. \ref{fig:comparison}).    We show furthermore that the QPI patterns obtained from such states, with generically weaker potentials to simulate out-of-plane native defects, agree much better with experiment than QPI maps obtained in previous theoretical calculations.

\begin{figure}
\includegraphics[width=1\linewidth,clip=true]{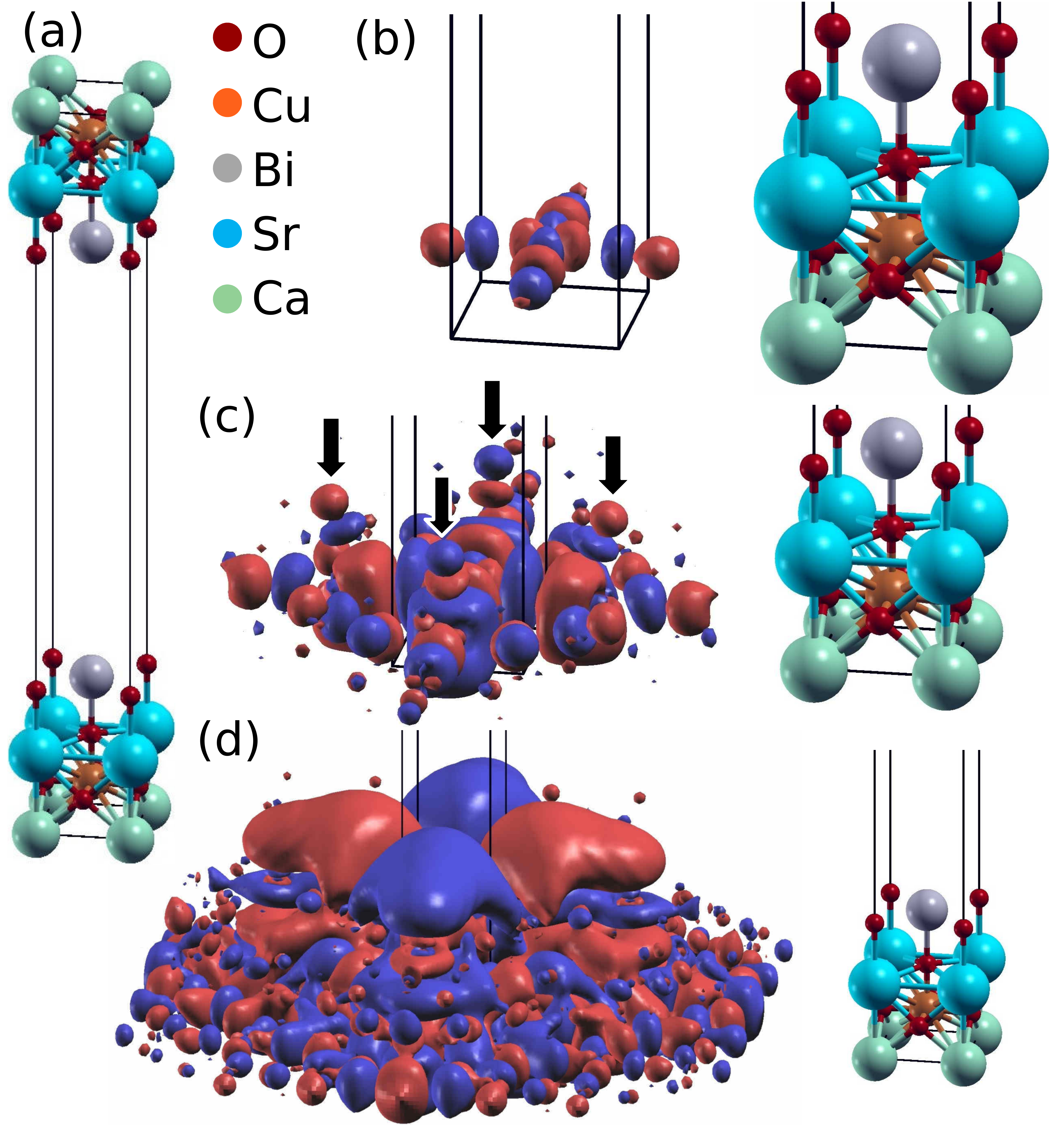}
\caption{
(color online) 
(a) Elementary cell used in first principles calculation to obtain the electronic structure
on the Bi$_2$Sr$_2$CaCu$_2$O$_8$ surface.
Isosurface plots of the Cu-$d_{x^2-y^2}$ Wannier function at 
(b) 0.05 bohr$^{-3/2}$, 
(c) 0.005 bohr$^{-3/2}$
and (d) 0.0002 bohr$^{-3/2}$.
Arrows indicate nearest-neighbor apical oxygen tails and 
red and blue indicate sign of the Wannier function.
}\label{fig:wannier}
\end{figure}
{\it Model.} The starting point of our investigation is first principles calculations of a BSCCO surface (Fig. \ref{fig:wannier}(a)) that yield
a one band tight-binding lattice model for the noninteracting electrons $c_{\mathbf{R}\sigma}$ (with Hamiltonian
$
H_0=\sum_{\mathbf{RR'},\sigma}t_{\mathbf{{\mathbf{RR}'}}}c_{\mathbf{R}\sigma}^{\dagger}c_{\mathbf{R'}\sigma}-\mu_0\sum_{\mathbf{R},\sigma}
c_{\mathbf{R}\sigma}^{\dagger}c_{\mathbf{R}\sigma}
\,,
 $
where  $t_{\mathbf{RR'}}$ are hopping elements between unit cells labeled $\mathbf R$ and  $\mathbf R'$
and $\mu_0$ is the chemical potential), and a Wannier basis
$w_{\R}(\rr)$ with $\rr$ describing the continuum position.
The Wannier function, obtained from a projected Wannier function analysis\cite{Ku_Wannier},  is primarily
of Cu-$d_{x^2-y^2}$ character with in-plane oxygen $p$-orbital contributions, as  can be seen in the isosurface plots for large values of the wave function, Fig.~\ref{fig:wannier}(b).
However, it also contains contributions from atomic wave functions in neighboring elementary cells, in particular those from the apical oxygen atoms above the Cu plane, Fig.  \ref{fig:wannier}(c). These are the main source of the large lobes above the neighboring Cu atoms at the position of the STM tip above the Bi-O plane, Fig. \ref{fig:wannier}(d). There is no weight, however, directly above the center Cu; see Fig. \ref{fig:wannier}(d). This can be understood from the fact that the hybridization of the Cu-$d_{x^2-y^2}$ orbital with apical O-$p$ and Bi-$p$ orbitals in the same unit cell is forbidden by symmetry.
In order to account for correlation effects at low energies, we use a mass renormalization factor
of $1/Z=3$ to scale down all hoppings such that the Fermi velocities approximately match the experimentally
observed values\cite{DamascelliRMP} and fix the chemical potential to be at optimal doping, ($n=0.85$).

\begin{figure}
\includegraphics[width=1\columnwidth]{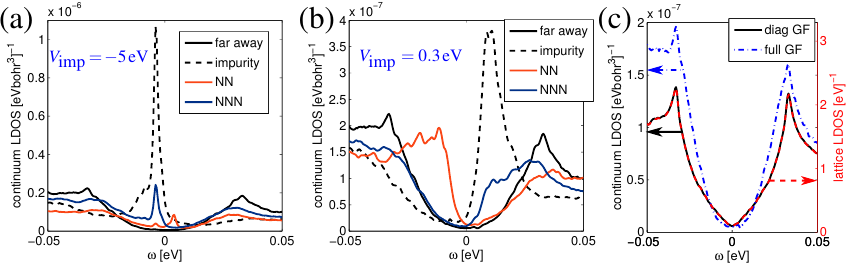}
\caption{(Color online) (a) Continuum LDOS at $5\text{\AA}$ above the BiO surface in the superconducting state with a single Zn impurity modeled by an onsite $-5\,\text{eV}$ potential. Shown are positions directly above Cu atoms far from the
impurity (black), at the impurity position (black, dashed), on the nearest neighbor
position (red [light gray]), and on the next-nearest neighbor position (blue [dark gray]), calculated using $20\times20$ supercells with broadening of $1\;\text{meV}$ and (b) for a weak impurity scatterer with $V_\text{imp}=0.3\,\text{eV}$ as used for the QPI analysis. 
In (c) we compare the spectral properties of the lattice density of states (red [light gray], dashed) with the continuum LDOS above a Cu atom calculated using the diagonal terms of the lattice Greens function $G_\sigma(\R,\R;\omega)$ only (black) and the full Greens function as given in Eq. (\ref{eq:continuumG}) (blue [gray], dash dotted); all of them calculated for a homogeneous superconductor and scale adjusted such that two curves (black and red [light gray], dashed) exactly overlay.
}
\label{fig:dos}
\end{figure}

Next, we solve the inhomogeneous mean field BdG equations for the full Hamiltonian of a superconductor in presence of an impurity
 $H=H_{0}+H_\text{BCS}+H_\text{imp},$
where the $d$-wave pairing interaction  $\Gamma_{\mathbf{RR'}}$ (details in the Supplemental Material
\cite{Note2})
enters the calculation of the  superconducting order parameter via  \mbox{$\Delta_{\mathbf{RR'}}=\Gamma_{\mathbf{RR'}}\langle c_{\mathbf{R'}\downarrow}c_{\mathbf{R}\uparrow}\rangle$} and gives rise to the second term
 $
 H_\text{BCS}=-\sum_{\mathbf{R}, \mathbf{R'}}\Delta_{\mathbf{RR'}}c_{\mathbf{R}\uparrow}^{\dagger}c_{\mathbf{R'}\downarrow}^{\dagger}+H.c.,
 $
while the third term is just a nonmagnetic impurity at lattice position   $\mathbf{R}^*$, e.g.
$
    H_\text{imp}=\sum_{\sigma}  V_\text{imp} c^\dagger_{\mathbf{R}^* \sigma}c_{\mathbf{R^*} \sigma}.
$
From the  BdG eigenvalues $E_{n\sigma}$ and eigenvectors $u_{n\sigma}$ and $v_{n\sigma}$ we can construct the usual retarded lattice Green's function
 \begin{equation}
\!\,G_\sigma(\R,\R';\omega)\!=\!\sum_n\! \left(\!\frac
{u^{n\sigma}_{\R} u^{n\sigma *}_{\R'}}{ \omega\!
-\!E_{n\sigma} \!
+\!i0^+ }\!
+\!
\frac{v^{n-\sigma}_{\R} v^{n-\sigma *}_{\R'}}{ \omega\!
+\!E_{n-\sigma} \!
+\!i0^+  }\!
\right)\!\!,\label{eq:latticeG}\!\!\!
\end{equation}
and the corresponding continuum Greens function\cite{Dellanna05,Choubey2014}
\begin{equation}
G_\sigma(\rr,\rr';\omega) =\sum_{\R,\R'}\!  G_\sigma(\R,\R';\omega) w_{\R}(\rr) w^*_{\R'}(\rr'),\label{eq:continuumG}
\end{equation}
by a simple basis transformation from the lattice operators $c_{\R\sigma}$ to the continuum operators $\psi_\sigma(\rr)=\sum_{\R} c_{\R\sigma}w_{\R}(\rr)$ where the Wannier functions $w_{\R}(\rr)$ are the matrix elements.
A similar transformation has been applied previously to understand neutron\cite{Walters2009} and x-ray\cite{Larson07,Abbamonte08} spectra in the normal state.
The continuum Green's functions can now be used to either calculate the LDOS $\rho(\rr,\omega)\equiv -\frac 1 \pi\mathop\text{Im}G_\sigma(\rr,\rr;\omega )$ as measured in STS experiments\cite{Tersoff1985} or obtain the QPI patterns by a Fourier transform.
Before considering an impurity, we note that the basis transformation in Eq. (\ref{eq:continuumG}) changes the spectral properties of the Greens function as it also contains terms that are nonlocal in the lattice description, e.g.  $G_\sigma(\R,\R';\omega)$ with $\R\neq \R'$. This has implications for the continuum LDOS $\rho(\rr,\omega)$, because the sign of $\mathop\text{Im}G_\sigma(\R,\R';\omega)$ is not fixed such that nonlocal contributions will lead to interference effects that can suppress or enhance the continuum LDOS at certain energies. These interference effects between Wannier functions are enhanced at the large distance from the surface where the STM tip is located and the Wannier functions are not confined by the lattice potential.
To illustrate this, we show in Fig. \ref{fig:dos}(c) the spectral dependence of the lattice LDOS for a homogeneous calculation which shows the well-known V-shape.
Applying the basis transformation by summing only over terms with $\R= \R'$, this behavior is not altered by the continuum LDOS as seen from the overlaied black curve, while in the full expression the spectral dependence is qualitatively modified and displays a clear U-shaped LDOS at low energies. Experimentally obtained conductances reveal exactly such a U-shaped behavior in overdoped samples\cite{Kohsaka08,McElroy05}, and the transition from V-shaped LDOS to more U-shaped has been observed with the same tip on samples with spatial inhomogeneous gaps\cite{Pan2001,Lang2002,Alldredge08}. We believe that these differences can be ascribed to the nonlocal contributions to Eq. (\ref{eq:continuumG}).
%
%

{\it Zn impurity. }  A Zn impurity substituting for Cu in BSCCO produces a strong attractive potential
which we simply model by an on-site potential of $V_\text{imp}=-5\,\text{eV}$, very similar to the value found in our first principles calculation (see Supplemental Material\cite{Note2}).
Calculating the LDOS, we find a sharp in-gap bound state peak around $\Omega_0=-3.6\,\text{meV}$, Fig.~\ref{fig:dos}(a).
The lattice LDOS from Eq.~(\ref{eq:latticeG}) shows a minimum at the impurity site and peaks at the NN sites [see Fig.~\ref{fig:comparison}(a) and Refs. \cite{BalatskyRMP,AlloulRMP}], precisely opposite from the experimental conductance map shown in Fig.~\ref{fig:comparison}(c).
As pointed out in Refs. \cite{Ting,abinitioZn,Martin,Markiewicz},
the problem lies in the consideration of the Cu lattice sites far from the BiO surface.
The correct quantity to study is the continuum LDOS
$\rho(\rr,\Omega_0)$
at the height of the STM tip, which
we assume to be at
$z=5\,\text{\AA}$ above the BiO surface.
The continuum LDOS obtained using Eq. (\ref{eq:continuumG}) presented in Fig.~\ref{fig:comparison}(b) indeed shows a maximum on the impurity site, originating from adding the NN apical oxygen tails of the Cu Wannier functions adjacent to the Zn site, and longer range intensity tails that are rotated 45 degrees from the nodal directions of the $d$-wave gap, in excellent agreement with the experimental observation as taken from Ref.[\onlinecite{PanZn}] Fig. \ref{fig:comparison}(c).
 We note a discrepancy on the 3rd site along the axis, where some of the reported experimental pattern are more intense than our theoretical result\cite{PanZn,Machida11,Hamidian12}. However, this peculiar feature seems not to be universal in experimental findings and might either be related to the local disorder environment on the surface of the crystal or the spatial supermodulation.
Finer resolution resonances reported in Ref. [\onlinecite{Hamidian12}] are also extremely similar to our calculations.
While this is crudely the same agreement reported by ``filter"-type theories\cite{Ting,Martin},  our calculation allows many further properties of the pattern to be recognized and provides a simple explanation of why they work.    
As in Ref. [\onlinecite{abinitioZn}], the theory allows us to compare the LDOS in the CuO$_2$ plane to that detected at the surface, but now also includes the redistribution of spectral weight (into, e.g., coherence peaks and impurity bound states) caused by the opening of the superconducting gap.

{\it QPI.}
\begin{figure}
\includegraphics[width=1\columnwidth]{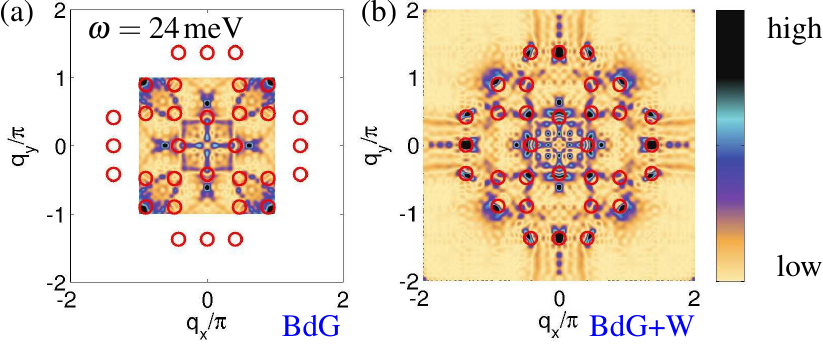}
\caption{(Color online) Simulated QPI pattern from conductance maps: (a) Fourier transform of the lattice LDOS (BdG) and (b) Fourier transform of the continuum LDOS (BdG+W) at the same energy $\omega=24\,\text{meV}$. Impurity potential for the weak scatterer $V_{\text{imp}}=0.3\,\text{eV}$. The red open symbols indicate the expected positions of the spots from the octet model.}
\label{fig_qpi_compare}
\end{figure}
QPI patterns in BSCCO are generated by several different types of disorder, believed to consist primarily of out-of-plane defects such as interstitial oxygens or site switching of Bi and Sr atoms, whose potentials are not known microscopically.
To account for these defects, we employ a weak potential scatterer on the Cu site with $V_\text{imp}=0.3\,\text{eV}$ and calculate the lattice LDOS and the continuum LDOS $\rho(\rr,\omega)$, both of which show only redistribution of spectral weight close to the impurity, compare Fig. \ref{fig:dos}(b).

Calculating the Fourier transform of the conductances $g(\rr,\omega)\propto \rho(\rr,\omega)$\cite{Hoffman2011} in order to obtain the conductance maps $|g({\mathbf q},\omega)|$ one is immediately faced with the problem that the lattice LDOS only contains information on length scales $\geq a$. Thus, the maps only extend in $\mathbf q$ space to the first Brillouin zone $[-\pi/a \ldots \pi/a]$, while the Fourier transform of the continuum LDOS is not  restricted in this way. The Fourier transformed maps have often  been analyzed in terms of the ``octet" model, which predicts a set of seven scattering vectors connecting hot spots at a given energy\cite{WangLee2003}. To compare to our result, we use the quasiparticle energies of our homogeneous superconductor to derive the expected QPI pattern.
Figure \ref{fig_qpi_compare} shows the calculated conductance maps $|g({\mathbf q},\omega)|$ at $\omega=24\,\text{meV}$ for (a) the lattice model (BdG) and (b) the Wannier method (BdG+W) where the $\mathbf q$ vectors from the octet model have been marked by circles. In the BdG-only result, a few of the spots are reproduced, others are absent
and more importantly, the large $\mathbf q$ vectors are not accessible with the lattice model.
In contrast, the map generated from the Wannier method shows a much better agreement with the octet model where all spots can be identified and no artificial spots appear.
A full scan of energies to visually highlight the dispersive features of the spots can be seen as an animation in the Supplemental Material\cite{Note2}.
Note that it is mathematically not possible to obtain the BdG+W maps from the corresponding BdG maps since the former also contain nonlocal contributions, as explained in Ref. [\onlinecite{Choubey2014}].

\begin{figure}
\includegraphics[width=1\columnwidth]{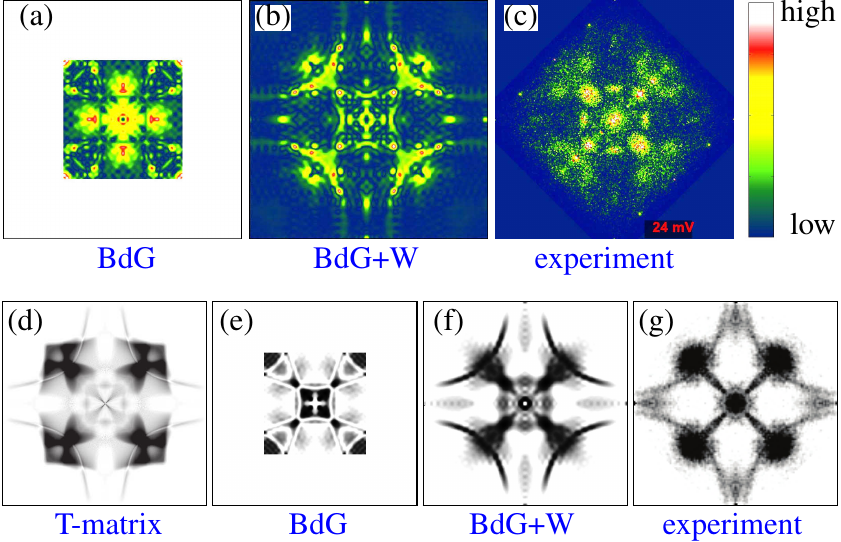}
\caption{(Color online)
(a) QPI Z map obtained from the Fourier transform of the real conventional space BdG patterns at $\omega=24\,\text{meV}$, (b) QPI Z map obtained from the Fourier transform of the continuous 3D LDOS$(x,y,z\approx 5\text{\AA};\omega)$ at $\omega=24\,\text{meV}$ showing strong similarity to (c) the experimental results reproduced from Ref. [\onlinecite{Fujita14}] and rotated to match the orientation in (a) and (b).
For the theoretical calculations a weak impurity scatterer of $V_{\text{imp}}=0.3\;\text{eV}$ was used.
(d) $T$-matrix scattering interference simulation for $\Lambda(q)$ from Ref. [\onlinecite{Fujita14}], (e) the same obtained from conventional BdG calculations, (f) $\Lambda$ map obtained from the Fourier transform of the continuous 3D LDOS showing strong similarity to the experimental results (g) reproduced from Ref. [\onlinecite{Fujita14}]. All $\bf q$ maps are from $[-2\pi/a \ldots 2\pi/a]$.}
\label{fig:comparison_int_qpi}
\end{figure}
In order to compare more closely to experiment, we follow Ref. [\onlinecite{Fujita14}] and
simulate the maps of the differential conductance ratios $Z({\mathbf q},\omega)$ as well as the energy integrated maps $\Lambda({\mathbf q})$ for both approaches, see definition in the Supplemental Material\cite{Note2}.
Figures \ref{fig:comparison_int_qpi}(a-c) show the Z maps of both methods side by side with an experimental result
\cite{Note1},
demonstrating the improvement of our method (BdG+W) compared to the  lattice BdG.
Similarly, we compare maps of the integrated ratio $\Lambda ({\mathbf{q}})$: \mbox{In Fig. \ref{fig:comparison_int_qpi}(g)} the experimental result is shown next to results from 3 different theoretical methods, (d) $T$-matrix simulation from Ref. [\onlinecite{Fujita14}], (e) lattice BdG and (f) our BdG-Wannier method. While all three theoretical models obtain large weight around $(\pm \pi,\pm \pi)$,  in agreement with experiment, only our Wannier method is capable of capturing simultaneously (1) the lines that extend from these large spots to the center, (2) the features along the axes between $\pi$ and $2\pi$ and (3) the arclike features that trace back the Fermi surface as in the analysis of Ref. \cite{Fujita14}.

{\it Conclusions.}  In this paper we have illustrated the utility of calculating the continuum rather than the lattice Green's function to compare with STM data in inhomogeneous systems, using
a first principles based Wannier function method.
We have focused on the cuprate superconductor BSCCO, and calculated the Zn resonant LDOS as well as QPI patterns, showing in both cases dramatic improvement compared to experiment relative to traditional lattice-based BdG analysis.
In the case of the Zn impurity, we have provided a first principles high-resolution theory of how electrons are transferred from nearest neighbor unit cells via apical oxygen atoms.
 In the case of the QPI patterns, the improved agreement is both with experiment and with the ``octet" model.  This shows that disagreements with the octet model in the past, primarily spurious arclike features and missing peaks, are due to the Fourier transform of the wrong electronic structure information: the lattice density of states in the CuO$_2$ plane rather than the continuous density of states at the STM tip position.
It is clear that our results have implications that go beyond the simple dispersing QPI patterns of a disordered BCS d-wave superconductor.  Any new theory of novel phenomena in the CuO$_2$ plane that seeks to compare with real space or QPI data should now be  ``dressed'' with Wannier information, or risk misidentification of crucial scattering features.

\textit{Acknowledgements.} The authors are grateful to J. Hoffman, Y. Wang, K. Fujita, and J. C. Davis for useful discussions.
P.C.,  A.K., and P.J.H. were  supported by DOE Grant No. DE-FG02-05ER46236, and T.B. as a Wigner Fellow at the Oak Ridge National Laboratory.
B.M.A. and A.K. acknowledge support from Lundbeckfond fellowship (grant A9318). W.K. was supported by DOE-BES Grant No. DE-AC02-98CH10886.
Work by T.B. was performed at the Center for Nanophase Materials Sciences, a DOE Office of Science user facility.

\clearpage
{\Large{[Supplemental Material]\\}}
In this supplementary material, we collect some technical details of our study,  exhibit some additional
results illustrating the parameter choices highlighted in
the main text, and compare theory and experiment over a wider range of parameters.
 \renewcommand{\thefigure}{\Roman{figure}}
  \renewcommand{\theequation}{S\arabic{equation}}
\setcounter{figure}{0}
\setcounter{equation}{0}

\textit{First principles based Wannier function calculations.}
The structural parameters of  Bi$_2$Sr$_2$CaCu$_2$O$_8$ were adopted  from Ref. \cite{Hybertsen1988}.
We applied the WIEN2K\cite{Blaha} implementation of the full potential linearized augmented plane wave
method in the local density approximation.
The band structure obtained in our work agrees well with that reported in Ref. \cite{Hybertsen1988}.
The calculation was performed for a body-centered tetragonal unit cell terminated at the BiO surface, with a slab of  approximately $18.5\,\text{\AA}$ vacuum.
      
\begin{figure}[tb]
  \includegraphics[width=0.7\columnwidth]{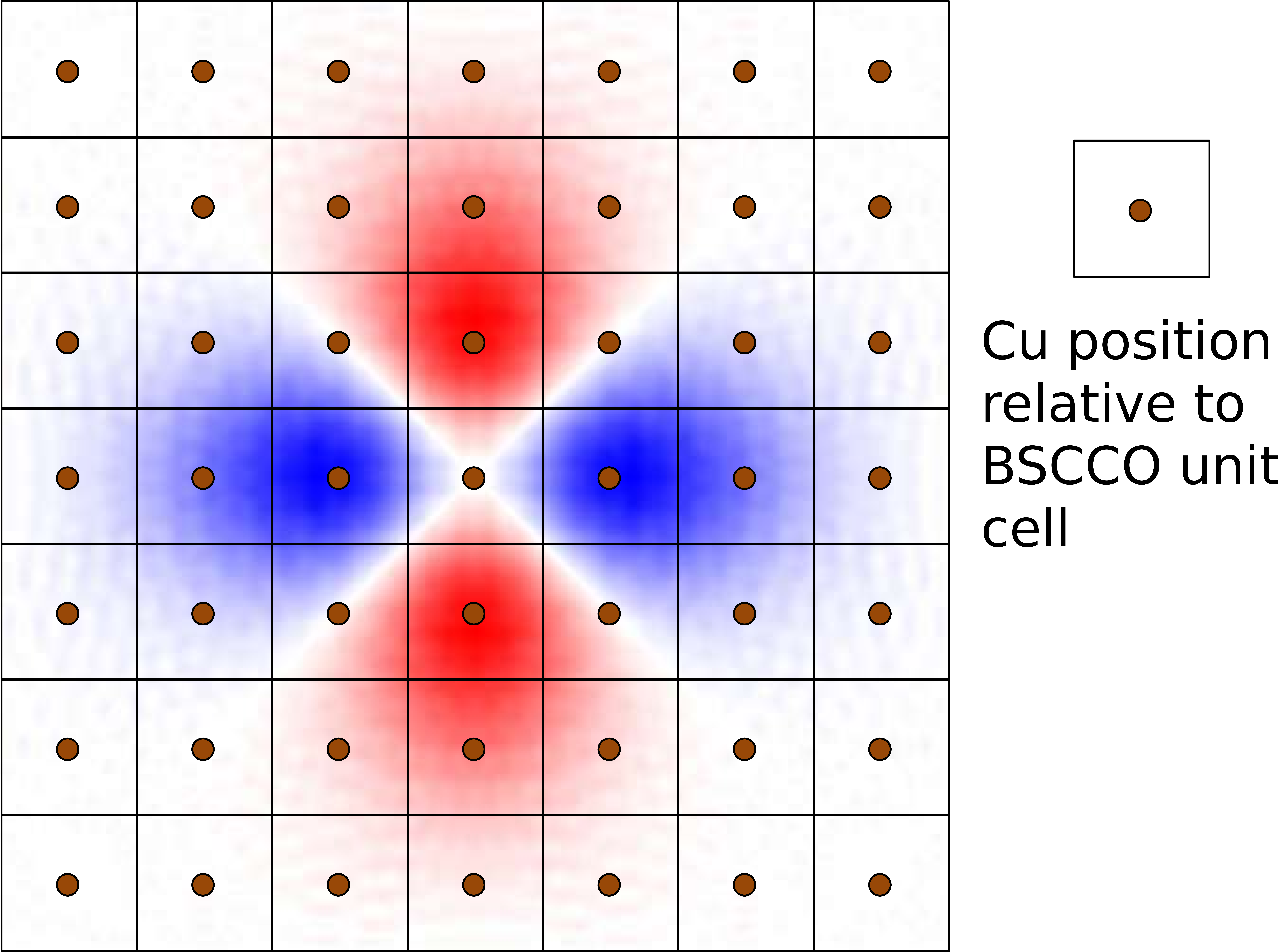}
 \caption{(Color online)
 Plot of the Cu-$d_{x^2-y^2}$ Wannier function at $5\,\text{\AA}$ above the BiO surface over a region of 7$\times$7 
 Bi$_2$Sr$_2$CaCu$_2$O$_8$ (BSCCO) elementary cells.
 }
 \label{fig_wannier}
 \end{figure}

Given the Bloch states $|{\bf k}j\rangle$ corresponding to a set of bands $\epsilon_{{\bf k}j}$, 
where ${\bf k}$ denotes the crystal momentum and $j$ denotes the band index, 
one can construct a set of Wannier states $|rn\rangle$ according to $|rn\rangle=\frac{1}{\sqrt{l}}\sum_{{\bf k}j} e^{-i{\bf k}\cdot {\bf R}} |{\bf k}j\rangle U_{jn}({\bf k})$,
where $l$ denotes the number of unit cells in the system,  ${\bf R}$ denotes the lattice vector and $n$ denotes the Wannier orbital index.
The matrix $U_{jn}({\bf k})$ fixes the so called gauge freedom of the Wannier functions.
For this purpose we use the projected Wannier function method \cite{Ku_Wannier,Anisimov2005} in which it is taken to be the projection of $n_{\rm{orb}}$ orbitals $|\varphi_{n} \rangle$ 
onto the Hilbert space of $n_{\rm{band}}(\geq n_{\rm{orb}})$ bands $|{\bf k}j\rangle$. 
Given that number of bands in the DFT calculation in this case exceeds the number of Wannier functions, the Wannier transformation is not gauge invariant. In a case like this the projected Wannier function method is preferred because it selects only the correct Hilbert space of the Cu-$d_{x^2y^2}$ symmetry. When using for example the maximally localization procedure\cite{Marzari97}, one risks localizing the Wannier functions at the expense of including other unwanted contributions in the Hilbert space.
The Wannier states are constructed by projecting the Cu-$d_{x^2-y^2}$ orbitals corresponding to the two Cu atoms in the unit cell shown in Fig. 2(a) of the manuscript on a [-3,3]eV window. To simplify the BdG calculation we reduce the resulting two band Hamiltonian to a one band Hamiltonian by cutting the out-of-plane hoppings, since they are an order of magnitude smaller than the in-plane hoppings.
The Cu-$d_{x^2-y2}$ Wannier function shown in Fig. 2 of the main text 
consists of a 141$\times$141$\times$67 real space grid centered at Cu and extends over 7$\times$7 unit cells.
In Fig. \ref{fig_wannier}, the same Wannier function is plotted at 5\AA \; above the BiO surface on a 501$\times$501 real space grid centered at Cu and extending over 7$\times$7 unit cells such that one can easily see the two features that give rise to the LDOS on the surface which deviates qualitatively from the one in the Cu-O plane: (1) there is no weight above the central Cu atom; (2) dominant weight is at the position of the nearest neighbor Cu atoms and further neighbors.

To calculate the potential of a Zn impurity in Bi$_8$Sr$_8$Ca$_4$Cu$_7$ZnO$_{32}$ supercell was used. 
The k-point mesh was taken to be 7$\times$7$\times$1 for the undoped normal cell 
and  4$\times$4$\times$1 for the supercell respectively.
The basis set sizes were determined by $RK_{\text{max}}=8$.
The on-site potential was found to be $-5.6\,\text{eV}$, similar to the value of $-5\,\text{eV}$ used in the calculations described. 

\textit{Tight binding model.}
\begin{figure}[tb]
 \includegraphics[width=0.85\columnwidth]{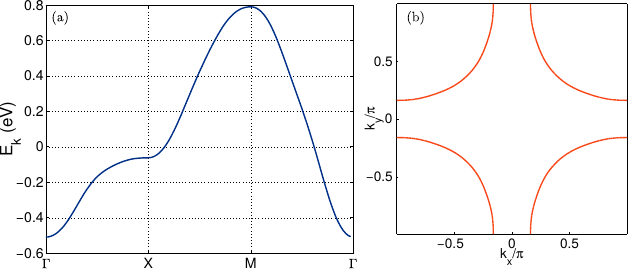}
\caption{(Color online) Band structure (a) and Fermi surface (b) of our one-band tight-binding model for \BSCCO at optimal doping.}
\label{fig_bands}
\end{figure}
\begin{table}
\caption{In-plane hoppings with descending magnitude that define our one-band model in Eq. (\ref{eq_tb}) for \BSCCO together with the connection vectors $\mathbf{R-R'}$.}
 \label{tab_1}
\renewcommand{\arraystretch}{1.4}
\begin{tabular*}{\columnwidth}{@{\extracolsep{\fill}}l|ccc}
\hline \hline
    &$\mathbf{R-R'}$  &$t_{\mathbf{R,R'}}$ (meV)  \\\hline
$t$ &  $(\pm 1,0,0)$, $(0,\pm 1,0)$       &-465.2        \\
$t'$ & $(\pm 1,\pm 1,0)$, $(\pm 1,\mp 1,0)$         &80.9         \\
 &  $(\pm 2,0,0)$, $(0,\pm 2,0)$       &-65.7        \\
  &  $(\pm 2,1,0)$, $(1,\pm 2,0)$,$(\pm 2,-1,0)$, $(-1,\pm 2,0)$       &-10.5        \\
 &  $(\pm 3,0,0)$, $(0,\pm 3,0)$       &-7.2        \\
 &  $(\pm 5,0,0)$, $(0,\pm 5,0)$       &4.5        \\
 &  $(\pm 4,0,0)$, $(0,\pm 4,0)$       &3.0        \\
  &  $(\pm 3,1,0)$, $(1,\pm 3,0)$,$(\pm 3,-1,0)$, $(-1,\pm 3,0)$       &-2.6        \\
  &  $(\pm 4,1,0)$, $(1,\pm 4,0)$,$(\pm 4,-1,0)$, $(-1,\pm 4,0)$       &1.0        \\
\hline \hline
\end{tabular*}
\end{table}
Our tight-binding model
\begin{equation}
H_0=\sum_{\mathbf{RR'},\sigma}t_{\mathbf{{\mathbf{RR}'}}}c_{\mathbf{R}\sigma}^{\dagger}c_{\mathbf{R'}\sigma}-\mu_0\sum_{\mathbf{R},\sigma}
c_{\mathbf{R}\sigma}^{\dagger}c_{\mathbf{R}\sigma}\label{eq_tb}
\,,
\end{equation}
describing the noninteracting electrons consists in a one band model with the hopping elements as shown in Table \ref{tab_1}. As mentioned in the main text, we fix the chemical potential ($\mu_0=-0.0160
\,\text{eV}$ without superconductivity) such that the Fermi surface in the normal state approximately matches the optimally doped case with a filling of $n=0.85$ electrons per spin and elementary cell.
The corresponding Fermi surface together with the band at a high symmetry cut, which has been renormalized by a factor of $Z = 3$ to account for correlations at low energies, are shown in Fig. \ref{fig_bands}.

\textit{Superconducting gap.}
In order to model a d-wave gap in our calculations, we simply take a repulsive nearest neighbor pairing interaction
with Hamiltonian
\begin{equation}
 H_{\text{BCS}}=-\sum_{\mathbf{RR'}} \Gamma_{\mathbf{RR'}}c_{\mathbf{R}\uparrow}^\dagger c_{\mathbf{R'}\downarrow}^\dagger c_{\mathbf{R'}\downarrow} c_{\mathbf{R}\uparrow}
\end{equation}
where  $\Gamma_{\mathbf{RR'}}=\Gamma_0$
for $\mathbf{R}$ and $\mathbf{R'}$ pointing to nearest neighbor elementary cells with $\Gamma_0=150\,\text{meV}$ such that the usual mean-field decoupling yields the Hamiltonian cited in the main text.
A self-consistent solution of the Bogoliubov-de Gennes (BdG) equations fixing the filling at $n=0.85$ for the homogeneous case in real space yields a converged gap and chemical potential which do not depend on the system size for calculations using more than $ N=30$ elementary cells.
We therefore use for all further calculations a lattice of 35$\times$35 elementary cells which converges to a $d$-wave gap of $\Delta_{\mathbf k}=\Delta_0/2 [\cos(k_x)-\cos(k_y)]$ with the constant $\Delta_0\approx 33\,\text{meV}$.

\begin{figure}[tb]
\includegraphics[width=0.32\columnwidth]{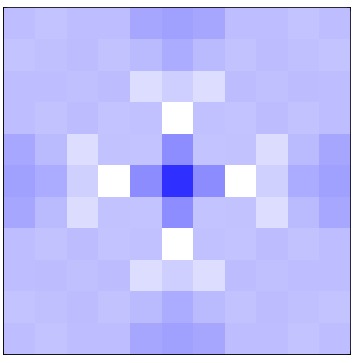}
 \rput[tr](-0.1\columnwidth,0.04\columnwidth){\setlength{\fboxsep}{0pt}\colorbox{white}{\color{red}$\omega=-30\,\text{meV}$}}
    \rput[tr](-0.27\columnwidth,0.31\columnwidth){\setlength{\fboxsep}{0pt}\colorbox{white}{(a)}}
\includegraphics[width=0.32\columnwidth]{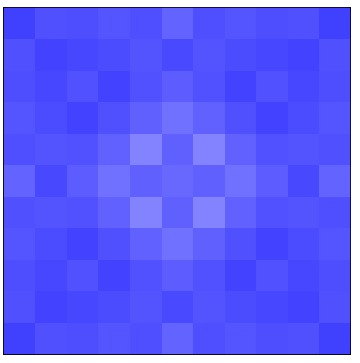}
 \rput[tr](-0.1\columnwidth,0.04\columnwidth){\setlength{\fboxsep}{0pt}\colorbox{white}{\color{red}$\omega=-18\,\text{meV}$}}
    \rput[tr](-0.27\columnwidth,0.31\columnwidth){\setlength{\fboxsep}{0pt}\colorbox{white}{(b)}}
\includegraphics[width=0.32\columnwidth]{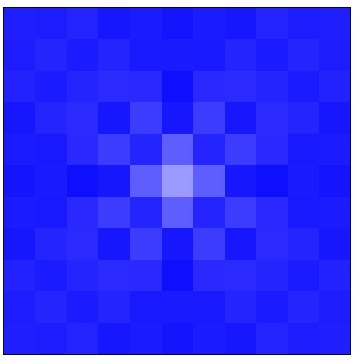}
\rput[tr](-0.1\columnwidth,0.04\columnwidth){\setlength{\fboxsep}{0pt}\colorbox{white}{\color{red}$\omega=-6\,\text{meV}$}}
    \rput[tr](-0.27\columnwidth,0.31\columnwidth){\setlength{\fboxsep}{0pt}\colorbox{white}{(c)}}\vspace{0.1cm}
\includegraphics[width=0.32\columnwidth]{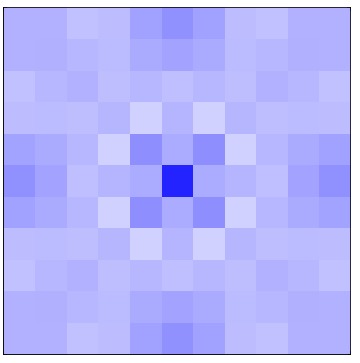}
\rput[tr](-0.1\columnwidth,0.04\columnwidth){\setlength{\fboxsep}{0pt}\colorbox{white}{\color{red}$\omega=30\,\text{meV}$}}
    \rput[tr](-0.27\columnwidth,0.31\columnwidth){\setlength{\fboxsep}{0pt}\colorbox{white}{(d)}}
\includegraphics[width=0.32\columnwidth]{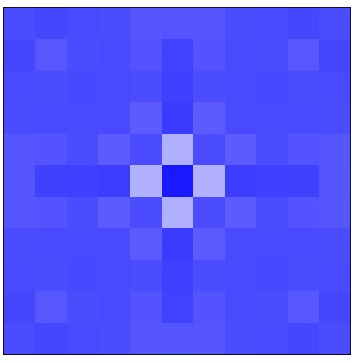}
\rput[tr](-0.1\columnwidth,0.04\columnwidth){\setlength{\fboxsep}{0pt}\colorbox{white}{\color{red}$\omega=18\,\text{meV}$}}
    \rput[tr](-0.27\columnwidth,0.31\columnwidth){\setlength{\fboxsep}{0pt}\colorbox{white}{(e)}}
\includegraphics[width=0.32\columnwidth]{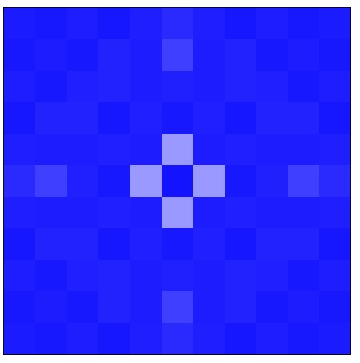}
    \rput[tr](-0.27\columnwidth,0.31\columnwidth){\setlength{\fboxsep}{0pt}\colorbox{white}{(f)}}
\rput[tr](-0.1\columnwidth,0.04\columnwidth){\setlength{\fboxsep}{0pt}\colorbox{white}{\color{red}$\omega=6\,\text{meV}$}}
\\[.2cm]
\includegraphics[width=0.48\columnwidth]{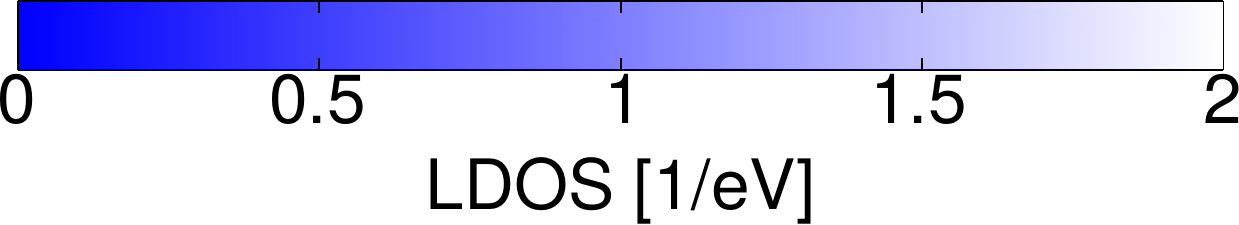}
\caption{(Color online)
Real space BdG patterns at $\omega=-30 \ldots 30\,\text{meV}$ as obtained from conventional BdG calculations with the weak scatterer for $V_{\text{imp}}=0.3\,\text{eV}$.
}
\label{fig_weak_scatt1}
\end{figure}

\begin{figure}[tb]
\flushleft\noindent
\includegraphics[width=0.32\columnwidth]{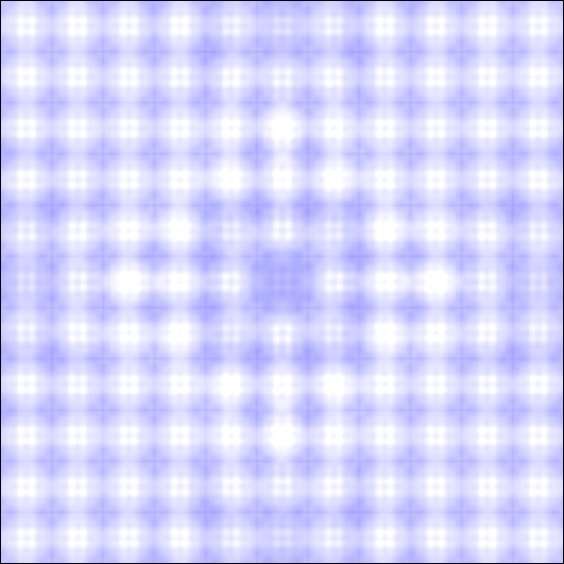}
 \rput[tr](-0.1\columnwidth,0.03\columnwidth){\setlength{\fboxsep}{0pt}\colorbox{white}{\color{red}$\omega=-30\,\text{meV}$}}
    \rput[tr](-0.27\columnwidth,0.31\columnwidth){\setlength{\fboxsep}{0pt}\colorbox{white}{(a)}}
\includegraphics[width=0.32\columnwidth]{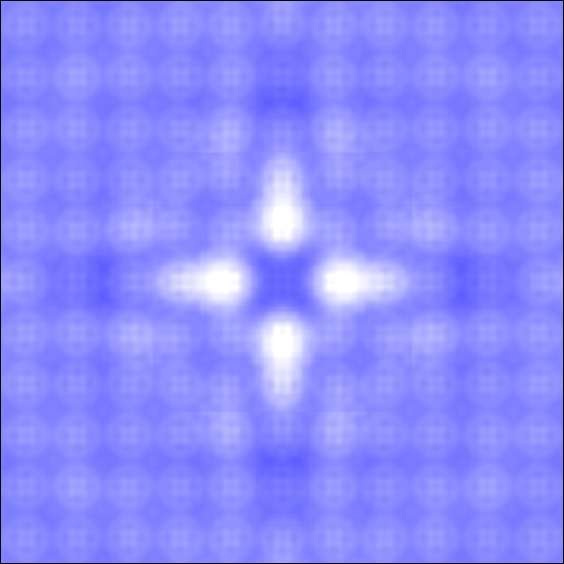}
 \rput[tr](-0.1\columnwidth,0.03\columnwidth){\setlength{\fboxsep}{0pt}\colorbox{white}{\color{red}$\omega=-18\,\text{meV}$}}
   \rput[tr](-0.27\columnwidth,0.31\columnwidth){\setlength{\fboxsep}{0pt}\colorbox{white}{(b)}}
\includegraphics[width=0.32\columnwidth]{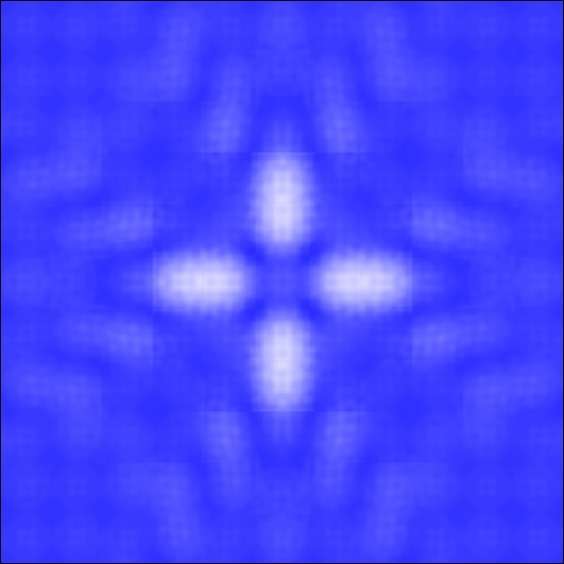}
\rput[tr](-0.1\columnwidth,0.03\columnwidth){\setlength{\fboxsep}{0pt}\colorbox{white}{\color{red}$\omega=-6\,\text{meV}$}}
   \rput[tr](-0.27\columnwidth,0.31\columnwidth){\setlength{\fboxsep}{0pt}\colorbox{white}{(c)}}\vspace{0.1cm}
\includegraphics[width=0.32\columnwidth]{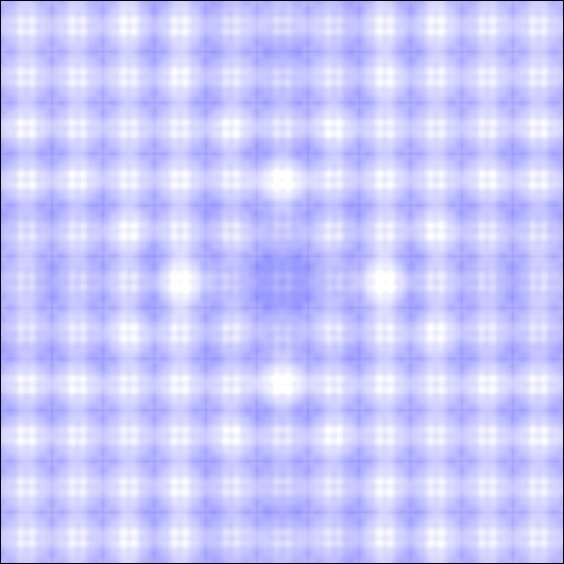}
\rput[tr](-0.1\columnwidth,0.03\columnwidth){\setlength{\fboxsep}{0pt}\colorbox{white}{\color{red}$\omega=30\,\text{meV}$}}
   \rput[tr](-0.27\columnwidth,0.31\columnwidth){\setlength{\fboxsep}{0pt}\colorbox{white}{(d)}}
\includegraphics[width=0.32\columnwidth]{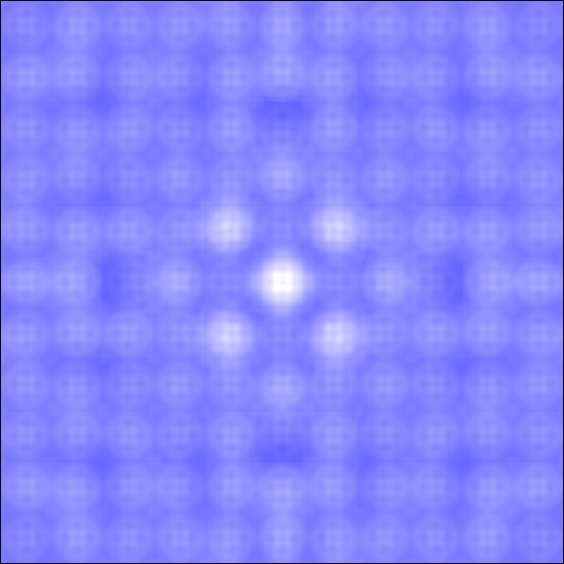}
\rput[tr](-0.1\columnwidth,0.03\columnwidth){\setlength{\fboxsep}{0pt}\colorbox{white}{\color{red}$\omega=18\,\text{meV}$}}
   \rput[tr](-0.27\columnwidth,0.31\columnwidth){\setlength{\fboxsep}{0pt}\colorbox{white}{(e)}}
\includegraphics[width=0.32\columnwidth]{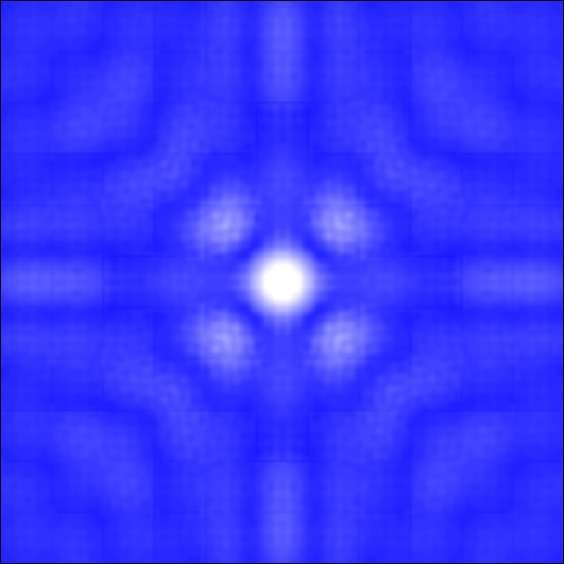}
\rput[tr](-0.1\columnwidth,0.03\columnwidth){\setlength{\fboxsep}{0pt}\colorbox{white}{\color{red}$\omega=6\,\text{meV}$}}
   \rput[tr](-0.27\columnwidth,0.31\columnwidth){\setlength{\fboxsep}{0pt}\colorbox{white}{(f)}}
\vspace{0.1cm}
\flushleft
\includegraphics[width=0.34\columnwidth]{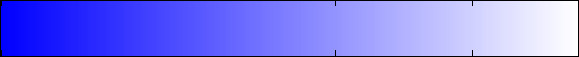}
\rput[tr](-0.34\columnwidth,-0.01\columnwidth){0}
\rput[tr](-0.145\columnwidth,-0.01\columnwidth){5}
\rput[tr](-0.065\columnwidth,-0.01\columnwidth){10}
\rput[tr](-0.0\columnwidth,-0.01\columnwidth){15}
\rput[tr](0.35\columnwidth,0.02\columnwidth){x $10^{-7} (\text{eV\,bohr}^3)^{-1}$}
\caption{(Color online)
Real space BdG+W patterns at $\omega=-30 \ldots 30\,\text{meV}$ as obtained using Eq. (3) of the main text for calculations with the weak scatterer for $V_{\text{imp}}=0.3\,\text{eV}$.
}

\label{fig_weak_scatt2}
\end{figure}

\textit{Lattice LDOS pattern for weak scatterer.}
As mentioned in the main text, the weak scatterer yields only small signatures of the impurity in the real space pattern which then turn into the pattern of the QPI maps when performing a Fourier transform.
For completeness, we show in Fig. \ref{fig_weak_scatt1} and \ref{fig_weak_scatt2} the analog of Fig. 1 (a) and (b) of the main text for $V_{\text{imp}}=0.3\,\text{eV}$. Experimentally, weak scatterers are present in a large number and of different types such that a comparison with their experimental signatures in real space is not possible.   For comparison to experimental data on BSCCO, this value of the potential was chosen simply because the QPI patterns resemble those seen in experiment best.    It is interesting to note however that this value does produce an in-gap bound state at around $10\,\text{meV}$, as seen in Fig. 3 of the main text, which corresponds roughly to the positions of the bound states seen for Ni impurities substituted for Cu\cite{Hudson2001}, and the real space patterns seen in Fig. \ref{fig_weak_scatt2} at nearby energies are extremely similar to that reported in experiment.
This comparison will be explored further in a future work.

\textit{Conductance maps in real space.}
\begin{figure}[tb]
\flushleft
\includegraphics[width=0.31\columnwidth]{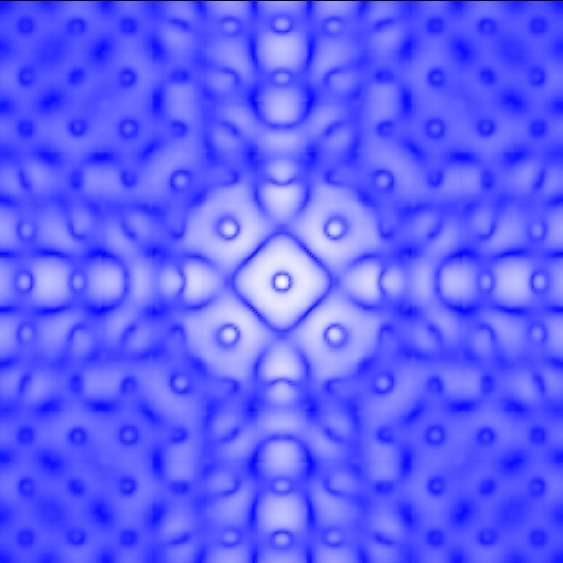}
 \rput[tr](-0.1\columnwidth,0.04\columnwidth){\setlength{\fboxsep}{0pt}\colorbox{white}{\color{red}$z_0=1.1\,\text{\AA}$}}
   \rput[tr](-0.27\columnwidth,0.31\columnwidth){\setlength{\fboxsep}{0pt}\colorbox{white}{(a)}}
 \vspace{0.05cm}
\includegraphics[width=0.31\columnwidth]{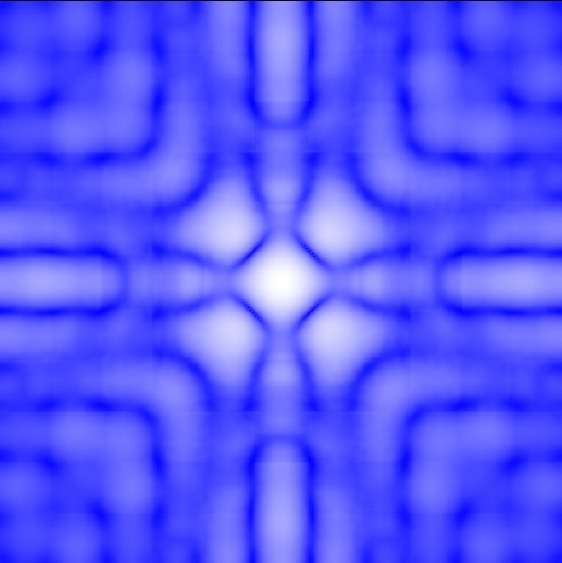}
 \rput[tr](-0.1\columnwidth,0.04\columnwidth){\setlength{\fboxsep}{0pt}\colorbox{white}{\color{red}$z_0=1.9\,\text{\AA}$}}
   \rput[tr](-0.27\columnwidth,0.31\columnwidth){\setlength{\fboxsep}{0pt}\colorbox{white}{(b)}}
 \vspace{0.05cm}
\includegraphics[width=0.31\columnwidth]{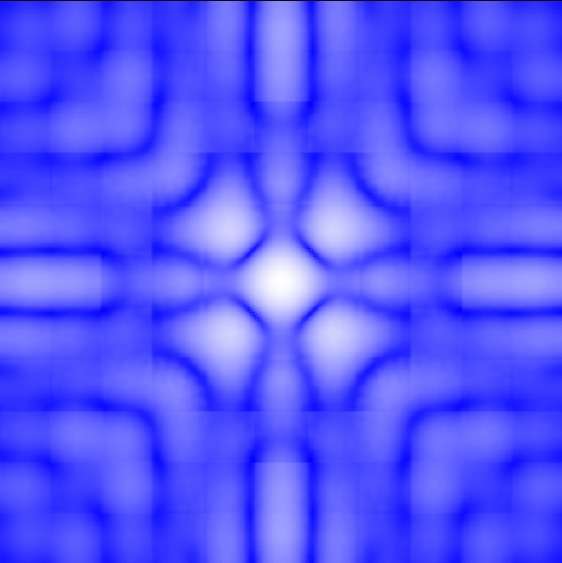}
 \rput[tr](-0.1\columnwidth,0.04\columnwidth){\setlength{\fboxsep}{0pt}\colorbox{white}{\color{red}$z_0=2.7\,\text{\AA}$}}
   \rput[tr](-0.27\columnwidth,0.31\columnwidth){\setlength{\fboxsep}{0pt}\colorbox{white}{(c)}}
\vspace{0.05cm}
\includegraphics[width=0.31\columnwidth]{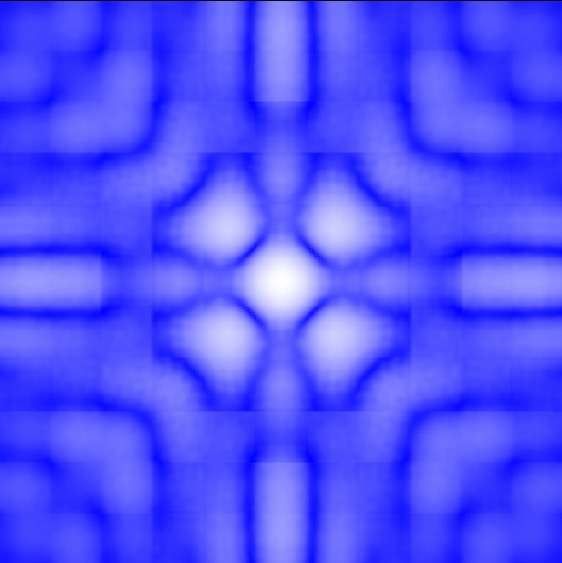}
\rput[tr](-0.1\columnwidth,0.04\columnwidth){\setlength{\fboxsep}{0pt}\colorbox{white}{\color{red}$z_0=3.4\,\text{\AA}$}}
   \rput[tr](-0.27\columnwidth,0.31\columnwidth){\setlength{\fboxsep}{0pt}\colorbox{white}{(d)}}
\vspace{0.05cm}
\includegraphics[width=0.31\columnwidth]{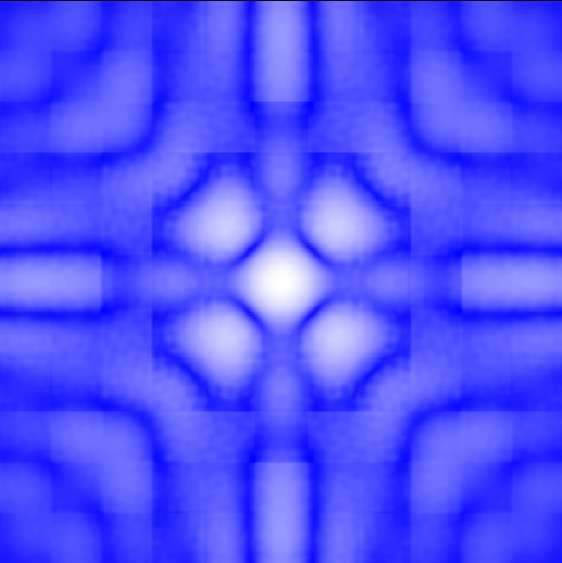}
\rput[tr](-0.1\columnwidth,0.04\columnwidth){\setlength{\fboxsep}{0pt}\colorbox{white}{\color{red}$z_0=4.2\,\text{\AA}$}}
   \rput[tr](-0.27\columnwidth,0.31\columnwidth){\setlength{\fboxsep}{0pt}\colorbox{white}{(e)}}
\vspace{0.05cm}
\includegraphics[width=0.31\columnwidth]{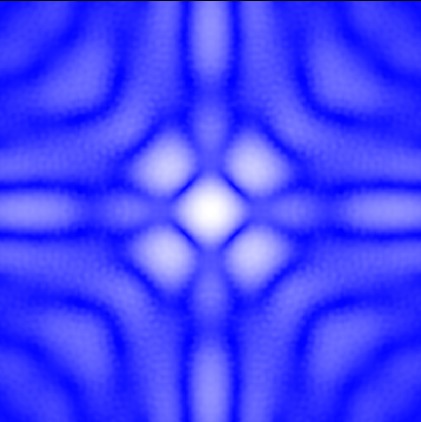}
\rput[tr](-0.1\columnwidth,0.04\columnwidth){\setlength{\fboxsep}{0pt}\colorbox{white}{\color{red}$z_0=5.0\,\text{\AA}$}}
   \rput[tr](-0.27\columnwidth,0.31\columnwidth){\setlength{\fboxsep}{0pt}\colorbox{white}{(f)}}
\\[0.2cm]
\centering
\includegraphics[width=0.5\columnwidth]{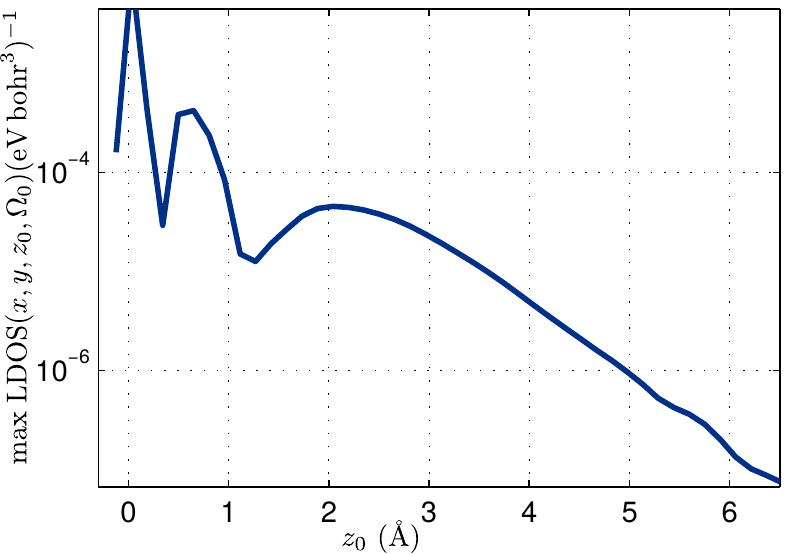}
\rput[tr](-0.52\columnwidth,0.34\columnwidth){\setlength{\fboxsep}{0pt}\colorbox{white}{(g)}}
 \includegraphics[height=0.35\columnwidth]{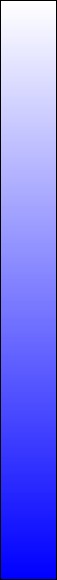}
  \rput[tr](0.07\linewidth,0.34\linewidth){high}
\rput[tr](0.07\linewidth,0.02\linewidth){low}
\caption{(Color online) (a-f)
Real space LDOS (BdG+W) patterns at $\Omega=-3.6\,\text{meV}$ for different heights $z_0$ above the Bi-O plane as obtained using Eq. (3) of the main text for calculations with the strong scatterer for $V_{\text{imp}}=-5\,\text{eV}$ plotted in logarithmic scale. The maximum of each plot is set to ``white'', and the numeric value can be read off from the graph (g).
}
\label{fig_strong_scatt1}
\end{figure}
\begin{figure}[tb]
\flushleft
\includegraphics[width=0.4\columnwidth]{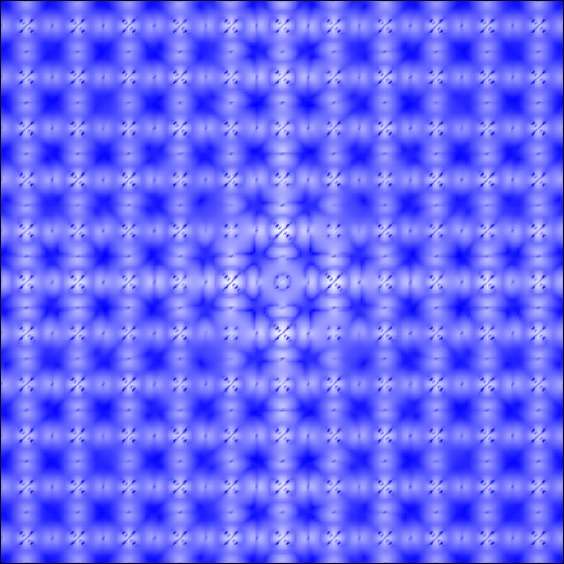}
 \rput[tr](-0.36\columnwidth,0.4\columnwidth){\setlength{\fboxsep}{0pt}\colorbox{white}{(a)}}
\vspace{0.1cm}
\includegraphics[width=0.4\columnwidth]{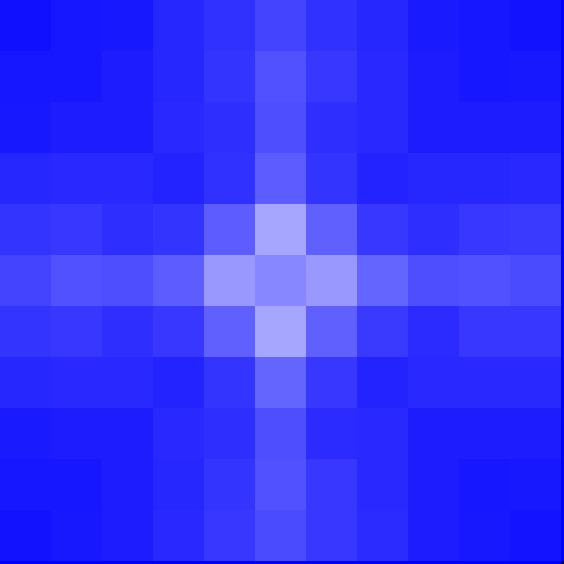}
 \rput[tr](-0.36\columnwidth,0.4\columnwidth){\setlength{\fboxsep}{0pt}\colorbox{white}{(b)}}
 \vspace{0.1cm}
  \includegraphics[height=0.35\columnwidth]{colorbar_bluemap.jpg}
  \rput[tr](0.07\linewidth,0.34\linewidth){high}
\rput[tr](0.07\linewidth,0.02\linewidth){low}
\caption{(Color online)
(a) Real space BdG+W pattern at  $\Omega=-3.6\,\text{meV}$ at the height of the Cu-O plane as obtained using Eq. (3) of the main text for calculations with the strong scatterer for $V_{\text{imp}}=-5\,\text{eV}$ plotted in logarithmic scale. The pattern on the right (b) additionally has been processed with a blur filter to visualize the resolution of one pixel per elementary cell.
}
\label{fig_strong_scatt2}
\end{figure}
The tunneling current in a STM experiment at a given bias voltage $V$ is given by\cite{Hoffman2011}
\begin{equation}
I(V,x,y,z)=-\frac{4\pi e}{\hbar} \rho_t(0) |M|^2 \int_0^{eV} \rho(x,y,z,\epsilon) d\epsilon\,,
\end{equation}
where $x,y,z$ are the coordinates of the tip, $\rho(x,y,z,\epsilon)$ is the continuum LDOS as defined in the main text, $\rho_t(0)$ is the density of states of the tip at the Fermi energy, and $|M|^2$ is the square of the the matrix element for the tunneling barrier.
 Note that the derivation of the above equation assumes an s-wave state in the STM tip\cite{Tersoff1985} which might not be true in the real experimental situation such that additional matrix element effects might modify the simple proportionality to the integrated LDOS. Taking the derivative with respect to the voltage yields the differential conductance
\begin{equation}
\frac{dI}{dV}=-\frac{4\pi e}{\hbar} \rho_t(0) |M|^2 \rho(x,y,z,eV)\,,
\label{eq_conductance}
\end{equation}
which is directly proportional to the local density of states a the tip position $\rho(x,y,z,eV)$ at energy $\omega=eV$. Note that the matrix element $|M|^2$ can in principle be calculated from the full information of the wave function of the sample, but also requires the knowledge of the wave function of the tip. For simplicity, we do not attempt to model the wave function of the tip and only work with the LDOS because by looking at relative differential conductance maps the matrix elements will drop out later.

The actual units in the corresponding figures of conductance maps are not shown in the main text, in the spirit of this proportionality (Eq. (\ref{eq_conductance})). Fig. \ref{fig_weak_scatt2} shows the maps for various energies at the same height including the full information about the units.
Fig. \ref{fig_strong_scatt1} shows the maps of the strong impurity scatterer as in Fig. 1 (b) of the main text, but for different heights above the BiO plane. Looking at the result, one sees that the LDOS rapidly enters the exponential limit as assumed in \cite{Tersoff1985} for deriving Eq. (\ref{eq_conductance}): The overall pattern is independent of the actual height, up to an overall scale change (b-f); note that Fig \ref{fig_strong_scatt1} (a) shows a map very close to the surface where Eq. (\ref{eq_conductance}) is not valid any more. In Fig. \ref{fig_strong_scatt2}, one again recognizes the pattern from the lattice LDOS at the Cu-O plane which is, of course, not the quantity measured in experiments.

\begin{figure}
\flushleft
\includegraphics[width=\columnwidth]{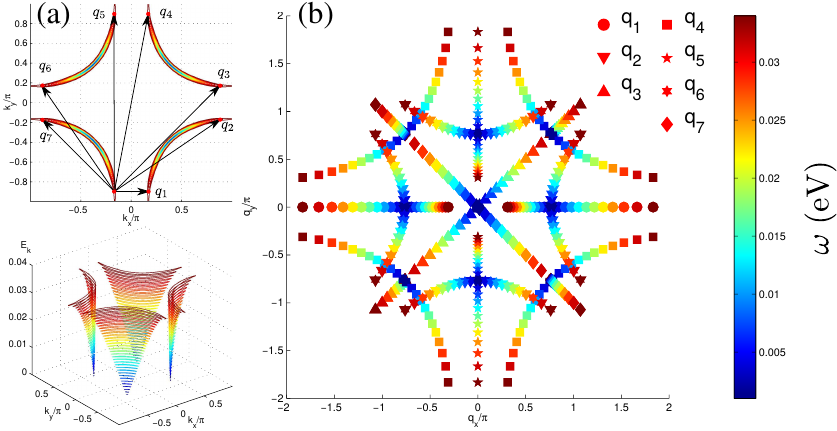}
\caption{(Color online)
(a) Isoenergy lines of the quasiparticle dispersion of the homogeneous d-wave superconductor as obtained from our model. The scattering vectors from one hot spot are indicated.
(b)
Expected dispersion of the spots according to the octet model.
}
\label{fig_octet}
\end{figure}

\textit{Fourier transforms of real space maps.}
\begin{figure}[tb]
\includegraphics[width=0.32\columnwidth]{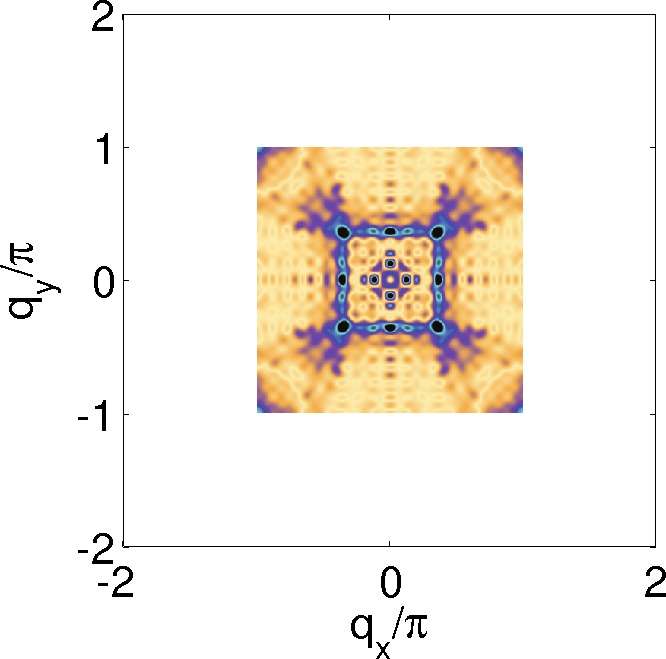}
 \rput[tr](-0.02\columnwidth,0.09\columnwidth){\setlength{\fboxsep}{0pt}\colorbox{white}{\color{red}$\omega=-30\,\text{meV}$}}
  \rput[tr](-0.22\columnwidth,0.30\columnwidth){\setlength{\fboxsep}{0pt}\colorbox{white}{(a)}}
\includegraphics[width=0.32\columnwidth]{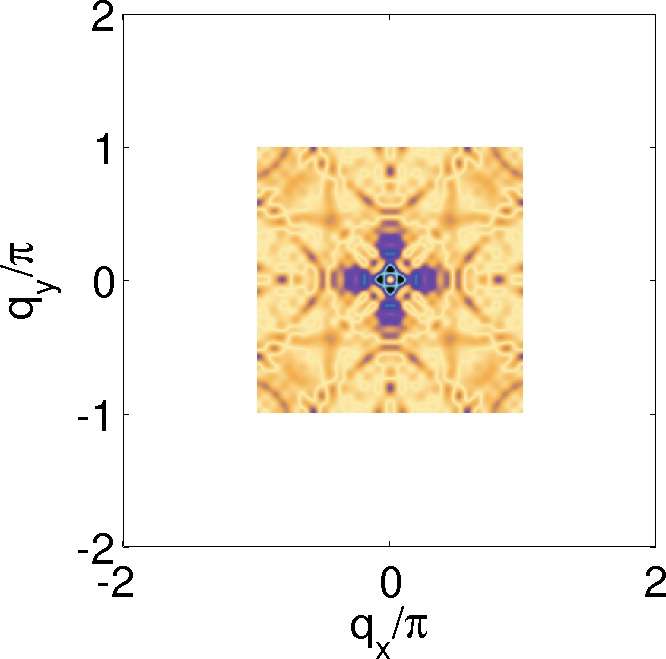}
 \rput[tr](-0.02\columnwidth,0.09\columnwidth){\setlength{\fboxsep}{0pt}\colorbox{white}{\color{red}$\omega=-18\,\text{meV}$}}
  \rput[tr](-0.22\columnwidth,0.30\columnwidth){\setlength{\fboxsep}{0pt}\colorbox{white}{(b)}}
\includegraphics[width=0.32\columnwidth]{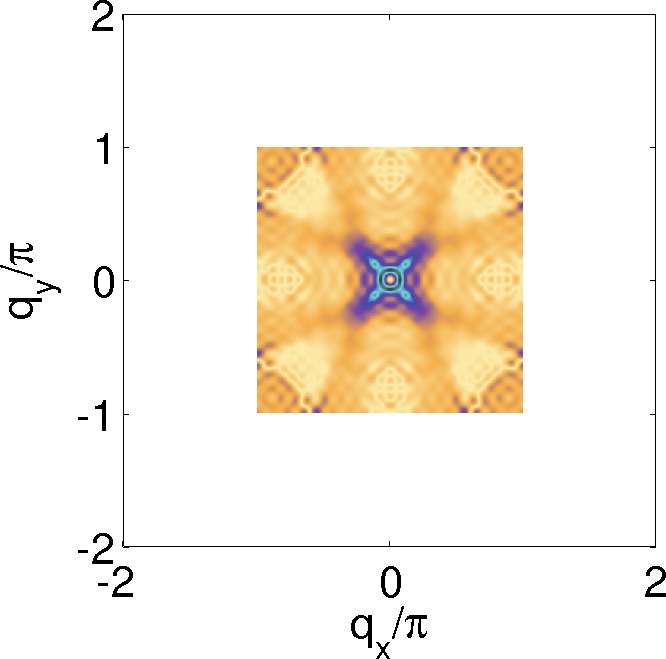}
\rput[tr](-0.02\columnwidth,0.09\columnwidth){\setlength{\fboxsep}{0pt}\colorbox{white}{\color{red}$\omega=-6\,\text{meV}$}}
  \rput[tr](-0.22\columnwidth,0.30\columnwidth){\setlength{\fboxsep}{0pt}\colorbox{white}{(c)}}\vspace{0.1cm}
\includegraphics[width=0.32\columnwidth]{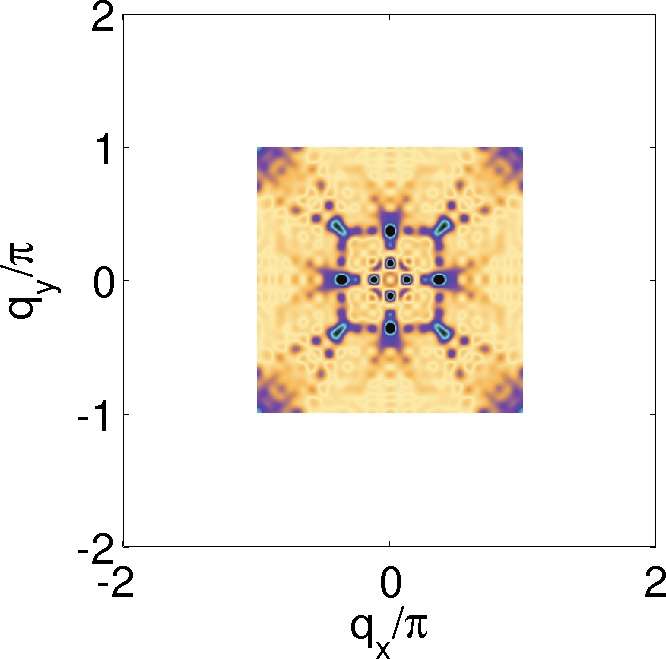}
\rput[tr](-0.02\columnwidth,0.09\columnwidth){\setlength{\fboxsep}{0pt}\colorbox{white}{\color{red}$\omega=30\,\text{meV}$}}
  \rput[tr](-0.22\columnwidth,0.30\columnwidth){\setlength{\fboxsep}{0pt}\colorbox{white}{(d)}}
\includegraphics[width=0.32\columnwidth]{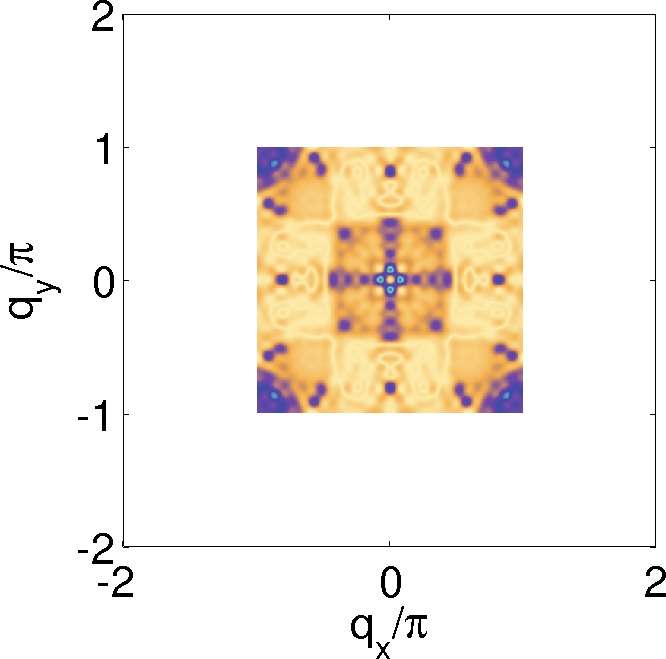}
\rput[tr](-0.02\columnwidth,0.09\columnwidth){\setlength{\fboxsep}{0pt}\colorbox{white}{\color{red}$\omega=18\,\text{meV}$}}
  \rput[tr](-0.22\columnwidth,0.30\columnwidth){\setlength{\fboxsep}{0pt}\colorbox{white}{(e)}}
\includegraphics[width=0.32\columnwidth]{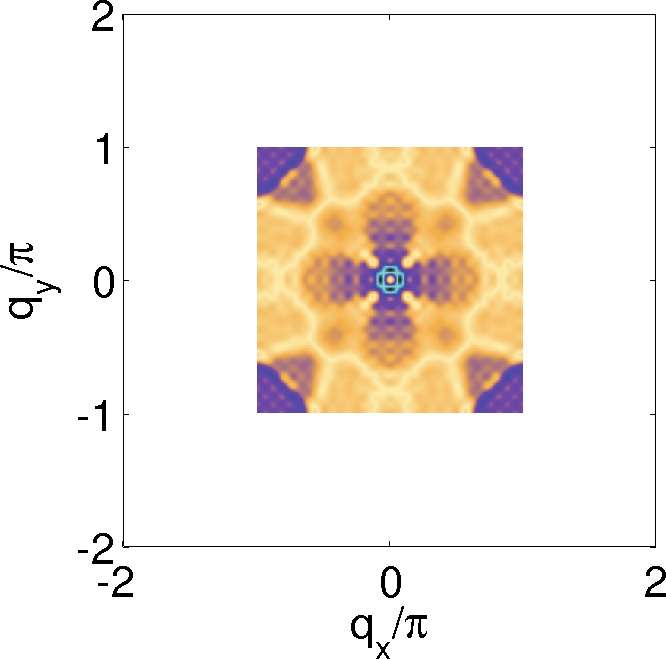}
\rput[tr](-0.02\columnwidth,0.09\columnwidth){\setlength{\fboxsep}{0pt}\colorbox{white}{\color{red}$\omega=6\,\text{meV}$}}
  \rput[tr](-0.22\columnwidth,0.30\columnwidth){\setlength{\fboxsep}{0pt}\colorbox{white}{(f)}}\vspace{0.3cm}
\includegraphics[width=0.32\columnwidth]{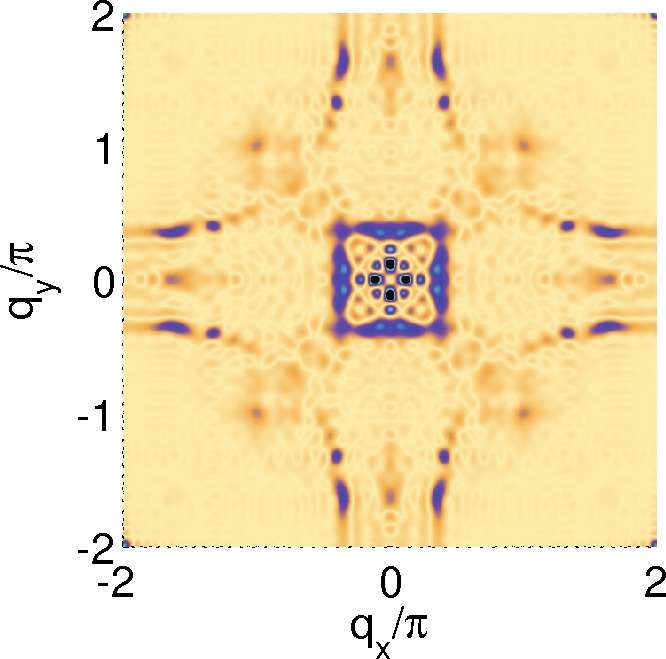}
 \rput[tr](-0.02\columnwidth,0.09\columnwidth){\setlength{\fboxsep}{0pt}\colorbox{white}{\color{red}$\omega=-30\,\text{meV}$}}
  \rput[tr](-0.22\columnwidth,0.30\columnwidth){\setlength{\fboxsep}{0pt}\colorbox{white}{(g)}}
\includegraphics[width=0.32\columnwidth]{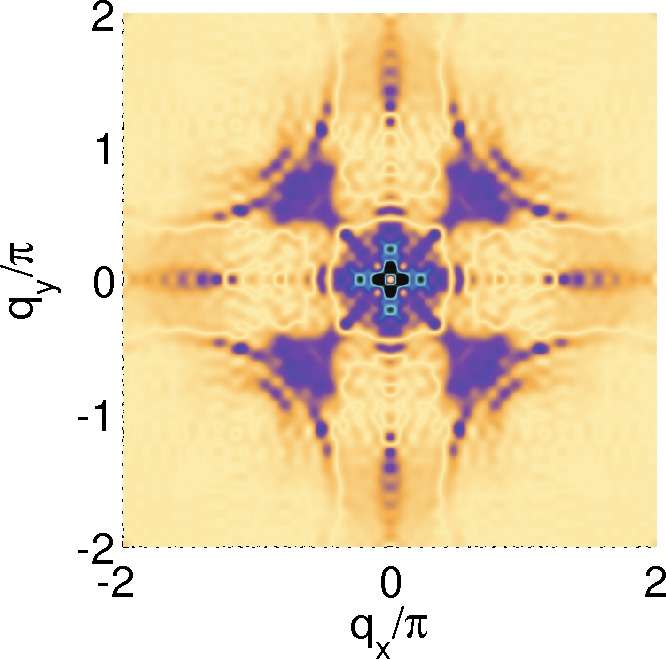}
 \rput[tr](-0.02\columnwidth,0.09\columnwidth){\setlength{\fboxsep}{0pt}\colorbox{white}{\color{red}$\omega=-18\,\text{meV}$}}
  \rput[tr](-0.22\columnwidth,0.30\columnwidth){\setlength{\fboxsep}{0pt}\colorbox{white}{(h)}}
\includegraphics[width=0.32\columnwidth]{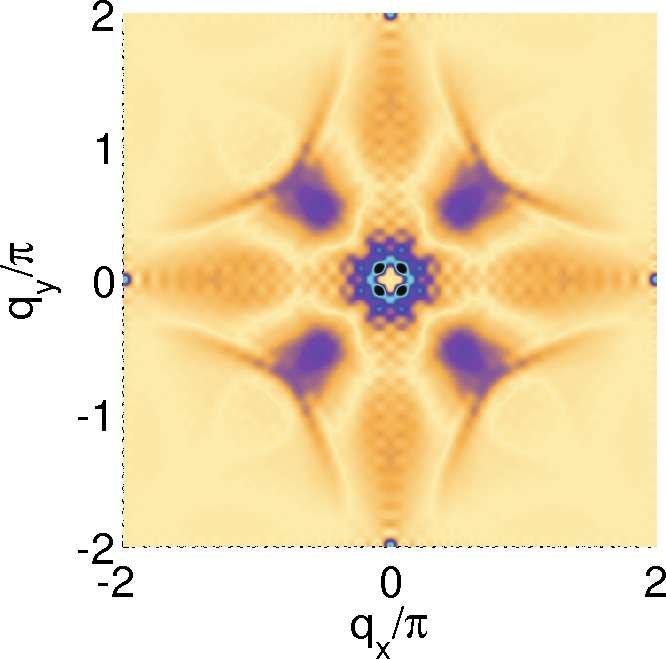}
\rput[tr](-0.02\columnwidth,0.09\columnwidth){\setlength{\fboxsep}{0pt}\colorbox{white}{\color{red}$\omega=-6\,\text{meV}$}}
  \rput[tr](-0.22\columnwidth,0.30\columnwidth){\setlength{\fboxsep}{0pt}\colorbox{white}{(i)}}\vspace{0.1cm}
\includegraphics[width=0.32\columnwidth]{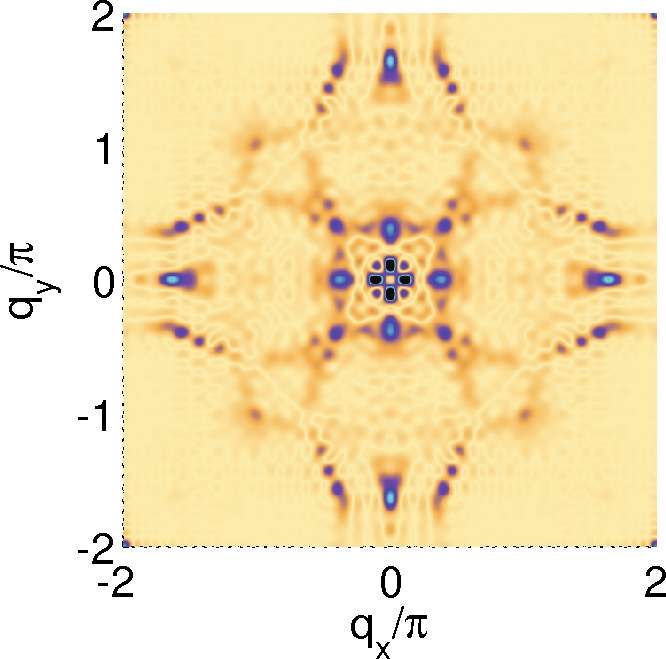}
\rput[tr](-0.02\columnwidth,0.09\columnwidth){\setlength{\fboxsep}{0pt}\colorbox{white}{\color{red}$\omega=30\,\text{meV}$}}
  \rput[tr](-0.22\columnwidth,0.30\columnwidth){\setlength{\fboxsep}{0pt}\colorbox{white}{(j)}}
\includegraphics[width=0.32\columnwidth]{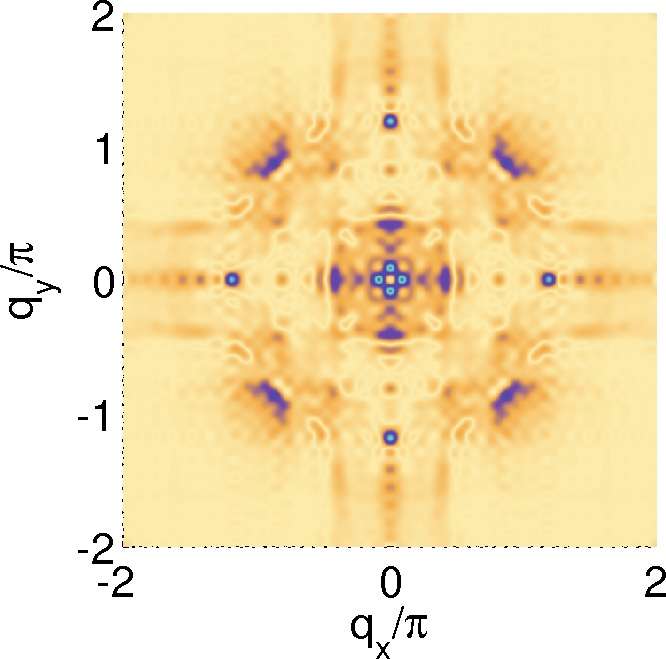}
\rput[tr](-0.02\columnwidth,0.09\columnwidth){\setlength{\fboxsep}{0pt}\colorbox{white}{\color{red}$\omega=18\,\text{meV}$}}
  \rput[tr](-0.22\columnwidth,0.30\columnwidth){\setlength{\fboxsep}{0pt}\colorbox{white}{(k)}}
\includegraphics[width=0.32\columnwidth]{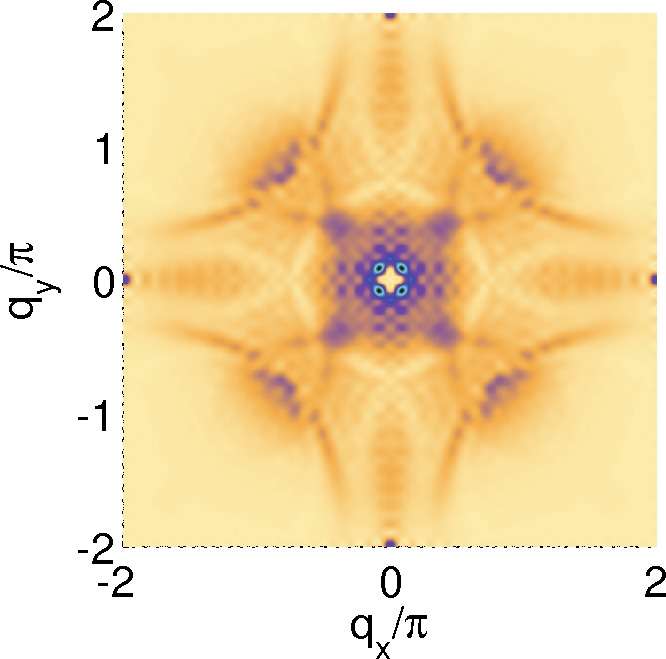}
\rput[tr](-0.02\columnwidth,0.09\columnwidth){\setlength{\fboxsep}{0pt}\colorbox{white}{\color{red}$\omega=6\,\text{meV}$}}
  \rput[tr](-0.22\columnwidth,0.30\columnwidth){\setlength{\fboxsep}{0pt}\colorbox{white}{(l)}}
\newline
\rotatebox{-90}{\includegraphics[width=0.035\columnwidth]{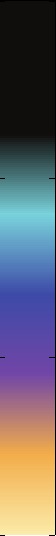}}
  \rput[tr](-0.4\linewidth,-0.01\linewidth){low}
\rput[tr](0.07\linewidth,-0.01\linewidth){high}
\caption{(Color online)
Fourier amplitudes $g(\mathbf{q},\omega)$ of the conductance maps from lattice BdG (a-f) and continuous LDOS maps (g-l).
Energies as indicated in the caption, impurity potential for the weak scatterer $V_{\text{imp}}=0.3\;\text{eV}$.
}
\label{fig_qpi_energy_g}
\end{figure}
Once the local density of states for the lattice model as well as the continuum maps at a given height are calculated, it is straightforward to perform a fast Fourier transform to obtain QPI maps. In order to obtain smooth maps in momentum space, we use a fine $\mathbf{k}$-mesh via the zero padding method. This is an interpolation scheme where a larger map in real space with data outside just filled with zeros is Fourier transformed. To avoid large oscillations from the central Bragg peak which will show up in the interpolated result, we remove it manually. A set of Fourier transformed maps of the lattice BdG result and the continuum LDOS (BdG+W) is shown in Fig. \ref{fig_qpi_energy_g} for a set of energies within the superconducting gap. Note that the position and the weight of the spots disperse with energy, and the BdG only result is restricted to the first Brillouin zone as explained in the main text.

\textit{Octet model.}
In order to compare our results to the octet model\cite{WangLee2003} which has been successfully used to describe the dispersive behavior of the spots measured in the QPI, we use our tight binding model together with the gap of the homogeneous system to calculate the 8 points  in the superconducting state corresponding to  the end-points of the banana-shaped isoenergy lines that are shown in Fig \ref{fig_octet} (a) for positive energies. The 7 connecting scattering vectors $\mathbf{q}_i(E)$ are defined the same figure, where in (b) the position of the spots (with the corresponding energy color-coded) is plotted to exhibit their dispersion in ${\bf q}$-space. The plot is restricted to positive energies, but the dispersive overall behavior is similar for negative energies, while the actual positions are slightly different because the band structure is not fully symmetric with respect to $E\leftrightarrow -E$.

\textit{Relative differential conductance maps and energy integrated maps.}
\begin{figure}
\includegraphics[width=0.32\columnwidth]{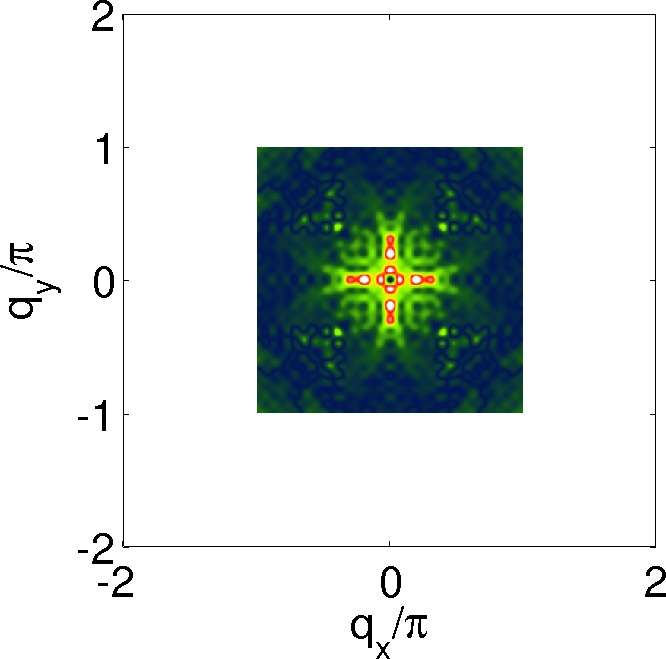}
 \rput[tr](-0.02\columnwidth,0.09\columnwidth){\setlength{\fboxsep}{0pt}\colorbox{white}{\color{red}$\omega=30\,\text{meV}$}}
  \rput[tr](-0.22\columnwidth,0.30\columnwidth){\setlength{\fboxsep}{0pt}\colorbox{white}{(a)}}
\includegraphics[width=0.32\columnwidth]{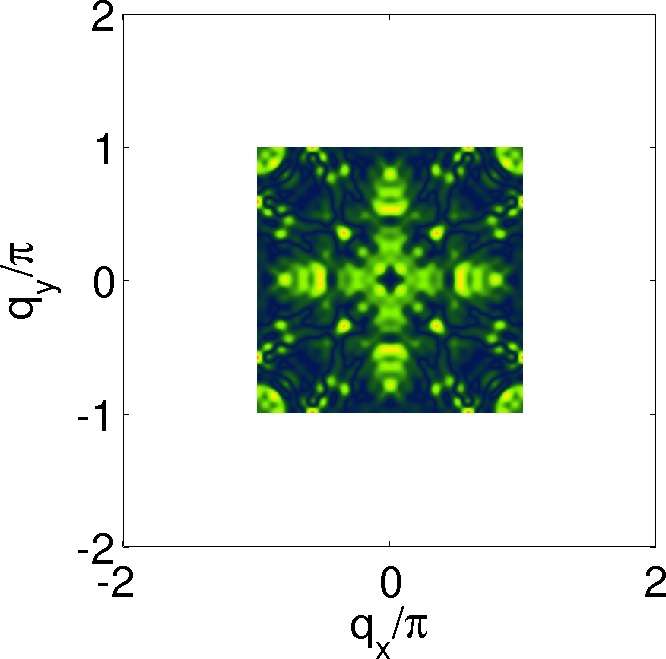}
 \rput[tr](-0.02\columnwidth,0.09\columnwidth){\setlength{\fboxsep}{0pt}\colorbox{white}{\color{red}$\omega=18\,\text{meV}$}}
  \rput[tr](-0.22\columnwidth,0.30\columnwidth){\setlength{\fboxsep}{0pt}\colorbox{white}{(b)}}
\includegraphics[width=0.32\columnwidth]{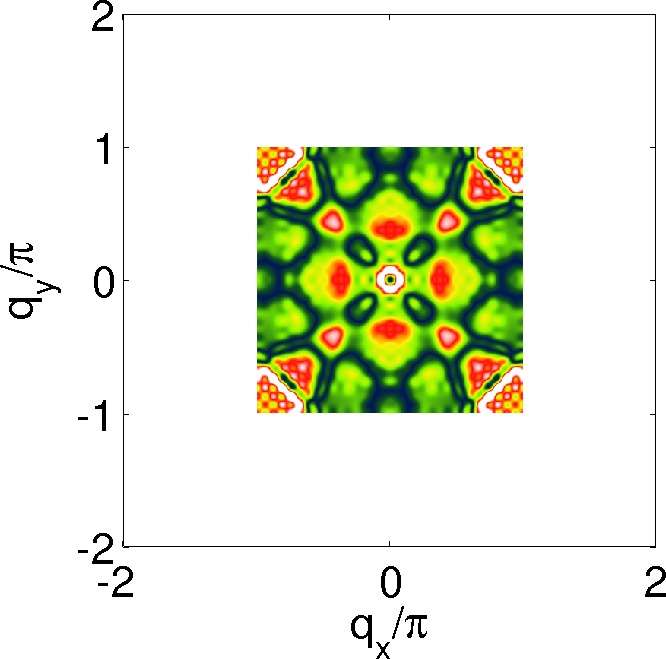}
\rput[tr](-0.02\columnwidth,0.09\columnwidth){\setlength{\fboxsep}{0pt}\colorbox{white}{\color{red}$\omega=6\,\text{meV}$}}
  \rput[tr](-0.22\columnwidth,0.30\columnwidth){\setlength{\fboxsep}{0pt}\colorbox{white}{(c)}}\vspace{0.3cm}
\includegraphics[width=0.32\columnwidth]{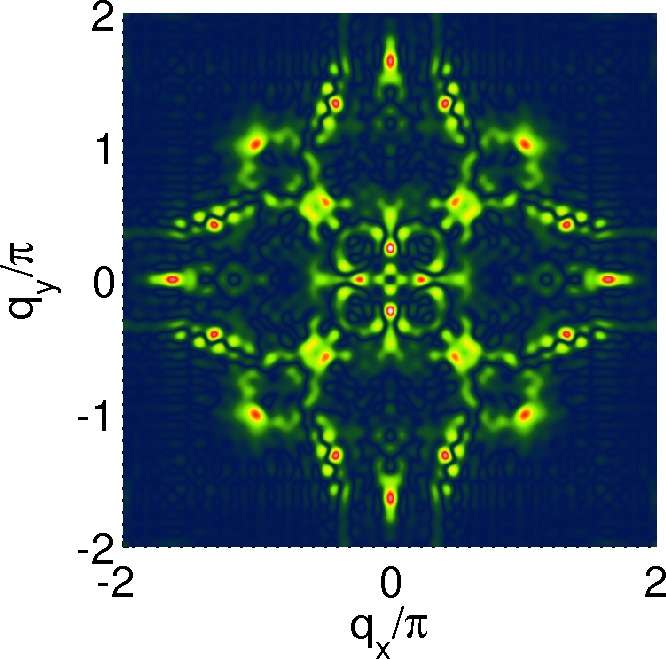}
 \rput[tr](-0.02\columnwidth,0.09\columnwidth){\setlength{\fboxsep}{0pt}\colorbox{white}{\color{red}$\omega=30\,\text{meV}$}}
  \rput[tr](-0.22\columnwidth,0.30\columnwidth){\setlength{\fboxsep}{0pt}\colorbox{white}{(d)}}
\includegraphics[width=0.32\columnwidth]{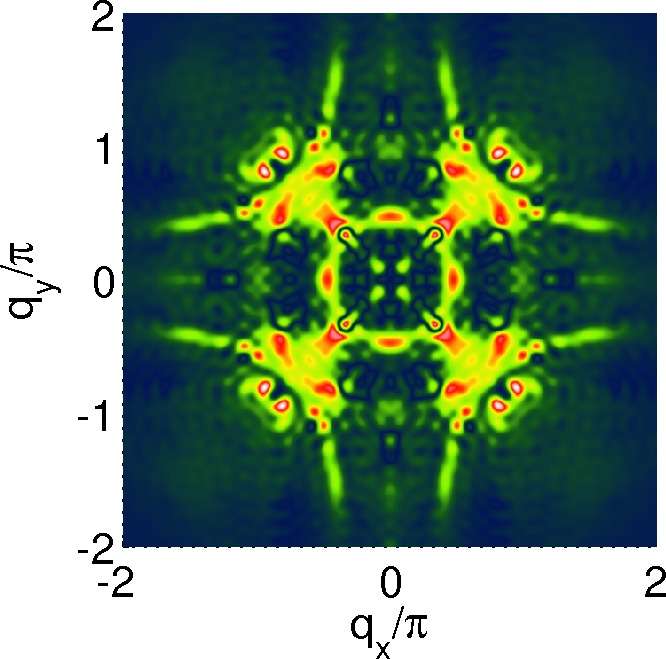}
 \rput[tr](-0.02\columnwidth,0.09\columnwidth){\setlength{\fboxsep}{0pt}\colorbox{white}{\color{red}$\omega=18\,\text{meV}$}}
  \rput[tr](-0.22\columnwidth,0.30\columnwidth){\setlength{\fboxsep}{0pt}\colorbox{white}{(e)}}
\includegraphics[width=0.32\columnwidth]{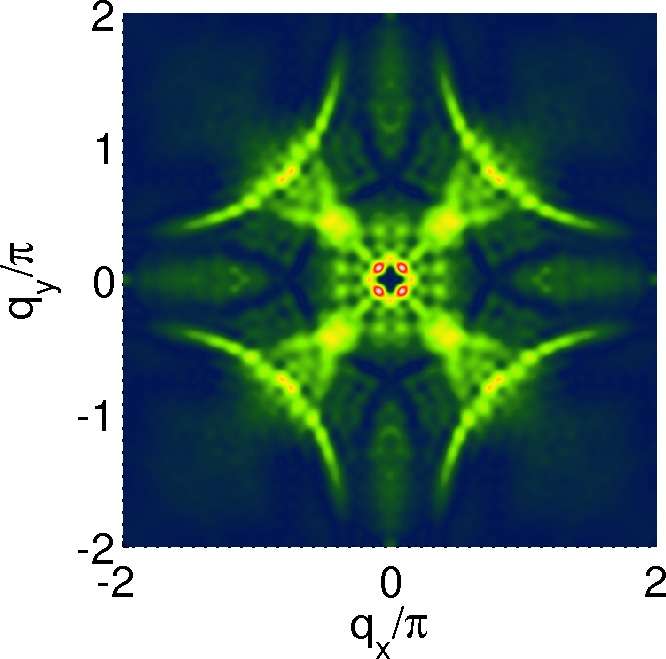}
\rput[tr](-0.02\columnwidth,0.09\columnwidth){\setlength{\fboxsep}{0pt}\colorbox{white}{\color{red}$\omega=6\,\text{meV}$}}
  \rput[tr](-0.22\columnwidth,0.30\columnwidth){\setlength{\fboxsep}{0pt}\colorbox{white}{(f)}}
\newline
\rotatebox{-90}{\includegraphics[width=0.035\columnwidth]{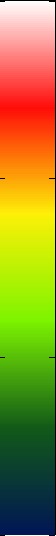}}
  \rput[tr](-0.4\linewidth,-0.01\linewidth){low}
\rput[tr](0.07\linewidth,-0.01\linewidth){high}
\caption{(Color online)
Fourier amplitudes of the relative conductance maps $Z(\mathbf{q},\omega)$ from lattice BdG (a-c) and continous ldos maps (BdG+W) (d-f).
Energies as indicated in the caption, impurity potential for the weak scatterer $V_{\text{imp}}=0.3\;\text{eV}$.
}
\label{fig_qpi_energy_z}
\end{figure}
\begin{figure}
 \frame{\includegraphics[width=0.33\columnwidth]{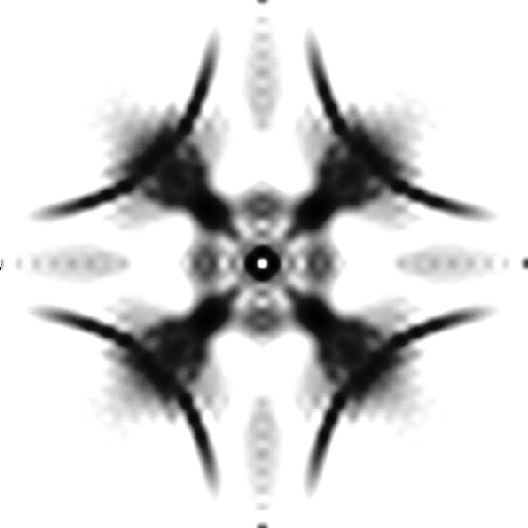}}
 \rput[tr](-0.00750\linewidth,0.3325\linewidth){
  \includegraphics[width=0.335\columnwidth]{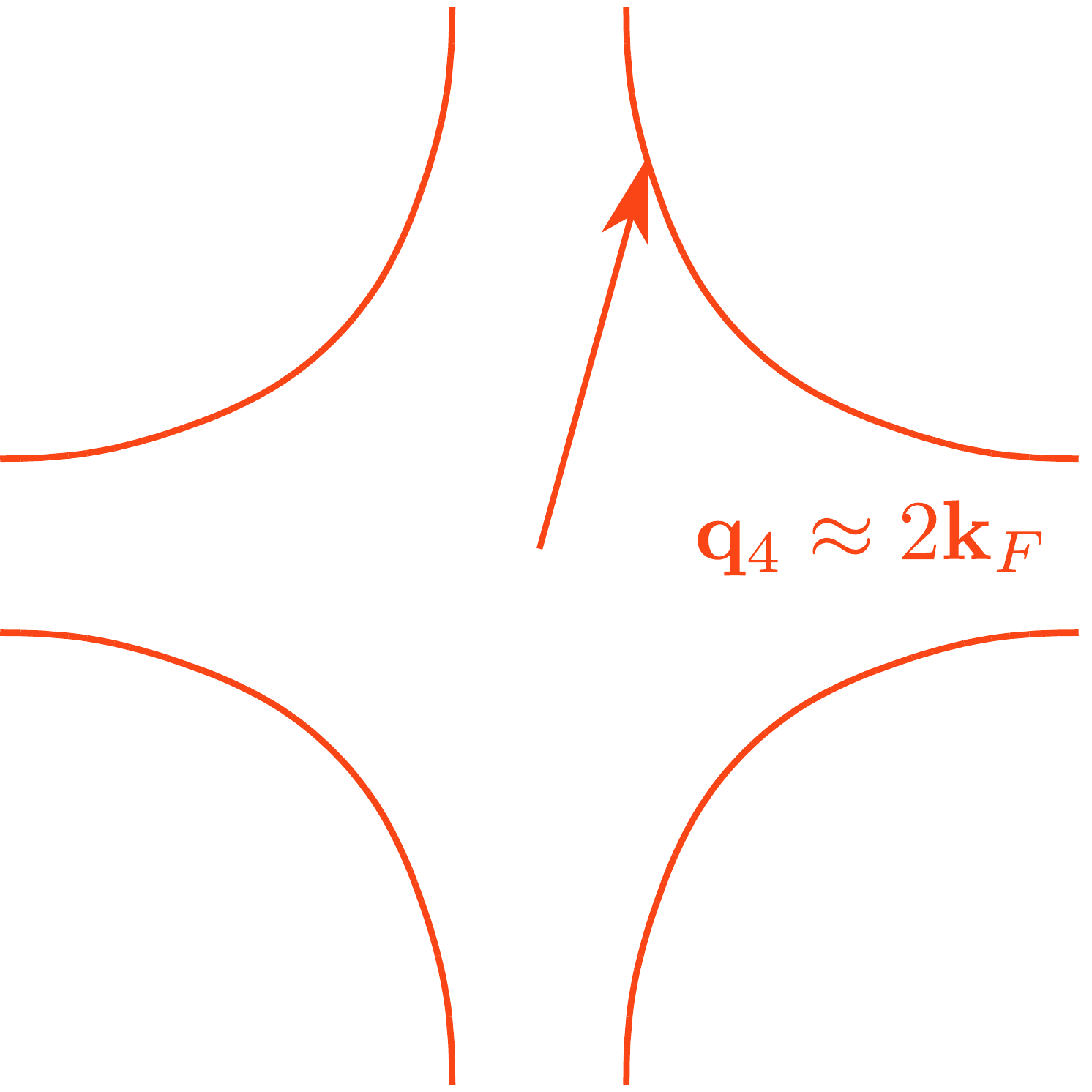}}
\caption{(Color online)
Integrated relative conductance maps $\Lambda(\mathbf q)$ as shown in the main text, but additionally with overlay of the Fermi surface from Fig. \ref{fig_bands} (b) scaled by a factor of two.
}
\label{fig_compare_int_FS}
\end{figure}
\begin{figure}
\animategraphics[controls,autoplay, palindrome, width=0.6\linewidth]{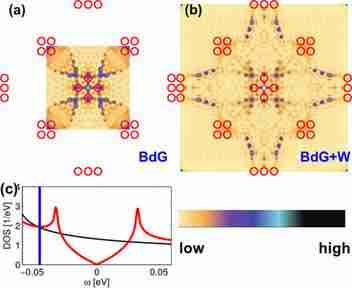}{./video_figs/}{1}{34}
\caption{(Color online)
Movie showing the energy evolution of the QPI pattern from conductance maps: (a) Fourier transform with zero padding of the lattice LDOS (BdG) and (b) Fourier transform with zero padding of the continuum LDOS (BdG+W), and (c) density of states of the homogeneous system together with a bar at the actual energy from $\omega=-50\,\text{meV}\ldots50\,\text{meV}$.  Impurity potential for the weak scatterer $V_{\text{imp}}=0.3\,\text{eV}$. The red open symbols indicate the expected positions of the spots from the octet model.
}
\label{fig_QPI_movie}
\end{figure}
As mentioned above, in the calculation of the conductivity, the matrix elements $|M|^2$ have not been taken into account due to lack of information about the wave function of the STM tip. Following e.g. Ref. \cite{Fujita14}, this complication can be avoided by taking the ratio of the differential conductances,
\begin{equation}
 Z(\rr,\omega)=\frac{g(\rr, \omega)}{g(\rr, -\omega)}=\frac{\rho(\rr, \omega)}{\rho(\rr, -\omega)}\,,\label{eq:integral}
\end{equation}
where the matrix elements simply drop out. In the Fourier transform of the ratio $Z(\mathbf{q},\omega)$ the overall structure of the spots becomes a bit more complicated, resulting in additional spurious peaks in the BdG case see Fig. \ref{fig_qpi_energy_z} (a-c) which are not observed in experiment. The BdG+W results for $Z(\mathbf{q},\omega)$ however compares quite reasonably with experimental results for energies corresponding to a  large fraction of the superconducting gap $\Delta_0$. Note that a direct comparison of the $Z(\mathbf{q},\omega)$ maps with the expected spots from the octet model is not possible since the maps contain information of the positive and negative energies where the scattering vectors are slightly different, $\mathbf{q}_i(E)\neq \mathbf{q}_i(-E)$. Recently, a method to trace back the Fermi surface by integrating over the Z maps in energy from zero to the superconducting gap has been introduced\cite{Fujita14,He2014}.   The quantity $\Lambda({\bf q})$ is defined as
\begin{equation}
 \Lambda({\mathbf{q}})=\int_0^{\Delta_0}\,d\omega\, Z(\mathbf{q },\omega)\,.
\end{equation}
Performing the integral using our Z maps for energies up to $\Delta_0$, we calculate the map presented in Fig. \ref{fig_compare_int_FS} and also in Fig. 5 (f) of the main text. The arc like features follow the Fermi surface blown up by a factor of two (orange line) because the scattering vector $\mathbf{q}_4$ follows the Fermi surface when sweeping in energy and the integral in Eq. (\ref{eq:integral}) accumulates weight along $2\mathbf{k}_F$, such that the method used in Refs. \cite{Fujita14,He2014} is demonstrated to work in our simulations.

\end{document}